\documentclass[pre,showpacs,preprint]{revtex4}

\usepackage{amsmath}

\usepackage{graphicx}
\usepackage{dcolumn}
\usepackage{bm}

\begin{document}

\title{Fluctuating Navier-Stokes equations for inelastic hard spheres or disks}
\author{J. Javier Brey, P. Maynar, and M.I. Garc\'{\i}a de Soria}
\affiliation{F\'{\i}sica Te\'{o}rica, Universidad de Sevilla,
Apartado de Correos 1065, E-41080, Sevilla, Spain}
\date{\today }

\begin{abstract}
Starting from the fluctuating Boltzmann equation for smooth inelastic hard spheres or disks, closed equations for the fluctuating hydrodynamic fields to Navier-Stokes order are derived. This requires to derive constitutive relations
for both the fluctuating fluxes and the correlations of the random forces. The former are identified as having the same form as the macroscopic average fluxes and involving the same transport coefficients. On the other hand, the
random force terms exhibit two peculiarities as compared with their elastic limit for molecular systems. Firstly,
they are not white, but have some finite relaxation time. Secondly, their amplitude is not determined by the macroscopic transport coefficients, but involves new coefficients.

\end{abstract}


\maketitle

\section{Introduction}
\label{s1}
Granular matter is ubiquitous in Nature. It has a tremendous impact on a wide range of industries and also raises important conceptual challenges. When the grains move freely and independently between collisions the system is referred to as a granular gas \cite{Ca90}. The simplest model of a granular gas at the particle level of description is an ensemble of inelastic hard spheres or disks \cite{BDyS97}. In the last decade or so, a great deal of progress and understanding of granular gases has been achieved \cite{Du00,Go03} by using this model, specially in the dilute limit in which the average behavior of the gas is accurately described by the inelastic Boltzmann equation \cite{GyS95,ByP04}. This includes derivation of the hydrodynamic Navier-Stokes equations with explicit expressions for the transport coefficients, study of several Newtonian and non-Newtonian steady states, analysis of the spectrum of the linearized inelastic Boltzmann operator, identification of several instabilities both in freely evolving and heated granular gases, and detailed characterization of the distribution function of several relevant states \cite{PyB03}. The theoretical predictions have been corroborated by molecular dynamics and direct Monte Carlo simulation results. On the other hand, knowledge about fluctuations and correlations in dilute granular gases is much more limited, although they are known to play a fundamental role in the macroscopic behavior of granular flows in many cases.

The purpose of this paper is to derive Langevin-like equations describing the dynamics of the fluctuating hydrodynamic fields in a freely evolving homogenous dilute granular gas, namely in the usually termed homogeneous cooling state (HCS) \cite{Ha83}. Of course, in the elastic limit the equations reduce to the fluctuating hydrodynamic equations proposed by Landau and Lifshitz for equilibrium molecular fluids \cite{LyL66}. The starting point will be the fluctuating Langevin-Boltzmann equation derived in ref. \cite{BMyG09} by using non-equilibrium statistical mechanics methods. The equation has the same mathematical form as the linearized inelastic Boltzmann equation plus and additive white noise term.

Hydrodynamic fluctuations in freely evolving granular gases have been already partially investigated. First studies used mesoscopic equations in which  the Landau and Lifschitz fluctuating hydrodynamic equations for molecular gases
were directly modified to incorporate the effect of energy dissipation in collisions, the resulting equations being expected to be valid in the quasi-elastic limit \cite{vNEByO97}. In particular, the correlations of the noise terms were assumed to be proportional to delta time functions (i.e. white noises) and determined by the Navier-Stokes transport coefficients, using the same expressions as for molecular systems. Also, some results based on a single relaxation model kinetic equation have been reported \cite{BMyR98}. In the above studies the interest focused on the initial buildup of spatial correlations of the hydrodynamic fields near the so-called clustering instability. Approaches starting from a more fundamental description of the granular gas, i.e. the Liouville equation of the system, have also been carried out. In particular, fluctuations and correlations of the total energy of an isolated granular gas have been computed and a good agreement between theory and simulations has been found \cite{BGMyR04,VPBvWyT06}. Moreover, in ref. \cite{BMyG09}, the fluctuations and correlations of the transversal component of the velocity field were studied in detail. It was shown that the Langevin equation obeyed by this field in a granular gas has two crucial differences with the elastic limit. Firstly, the noise is not white but has a finite correlation time and, secondly, the amplitude of its second moment is not determined by the shear viscosity but by some new coefficient. It must be noticed that the transversal velocity field is special in the sense that it evolves uncoupled from the other hydrodynamic fields.

The purpose of this paper is to extend the analysis in \cite{BMyG09}, by deriving the whole closed set of fluctuating hydrodynamic equations for a dilute granular gas in the HCS. In this context, it must be realized that this state is the homogeneous reference state for granular gases, playing a role somehow similar to the equilibrium state in molecular fluids. Therefore, understanding fluctuations and correlations in this state is an unavoidable first step
towards a deep knowledge of  them in more complex and also more realistic states of granular gases.

It is worth to mention also some previous studies of fluctuations dealing with driven granular gas models, in which the grains are assumed to be in contact with some external energy source or thermal bath \cite{PByL02,PByV07,MGyT09}.
Since the latter modifies the stochastic properties of the granular gas in a nontrivial way, it is not evident that there is a direct relation between fluctuations in driven models and the free model being considered here.

The plan of the remaining of the paper is as follows. In Sec.\ \ref{s2}, the Boltzmann-Langevin equation for smooth inelastic hard spheres or disks is shortly reviewed, and the appropriate dimensionless time and length scales are introduced. From the above kinetic equation, balance equations for the fluctuating hydrodynamic fields in the HCS are directly derived, as shown in Sec.\ \ref{s3}. These equations are worthless until they are closed by deriving constitutive relations for the fluxes appearing in them and also explicit expressions for the correlations of the noise terms.  This is done here by introducing a projection operator over the hydrodynamic subspace of the distribution functions. This subspace is generated by the hydrodynamic eigenfunctions of the linear inelastic Boltzmann operator, whose definitions and expressions are also briefly reminded. The equation for the number density involves neither flux nor noise term. The Langevin equation  for the velocity field is derived in Sec.\ \ref{s4} and the one for the energy field in Sec.\ \ref{s5}.  In both cases, expressions for the fluctuating fluxes and for the correlation functions of the fluctuating forces are provided. These expressions are not exact, but have been obtained under well defined and controlled approximations. It is shown that the fluctuating forces have finite relaxation times which can be related with non-hydrodynamic eigenvalues of the linearized Boltzmann operator. For the sake of clarity, many details of the calculations as well as technical points are given in six appendixes. Finally, Sec. \ref{s6} contains a short summary and some final remarks.

\section{Boltzmann-Langevin equation for smooth inelastic hard spheres or disks}
\label{s2}
The system being considered is a dilute gas of smooth inelastic hard spheres ($d=3$) or disks
($d=2$) of mass $m$ and diameter $\sigma$. Inelasticity of collisions is modeled by means
of a constant, velocity independent, coefficient of normal restitution $\alpha$, defined in the
interval $0 < \alpha \leq 1$. At a mesoscopic level, the system is described by a fluctuating
or stochastic one-particle distribution function, $F({\bm r}, {\bm v},t)$, whose average, denoted by
$f({\bm r},{\bm v},t)$, is the usual one-particle distribution function giving the average number of particles
density being at position ${\bm r}$ with velocity ${\bm v}$ at time $t$. This average distribution obeys
the nonlinear inelastic Boltzmann equation \cite{GyS95}.

The system is assumed to be in the homogeneous cooling state (HCS), macroscopically characterized by a uniform
number of particles density
$n$, a vanishing flow velocity, and a uniform granular temperature $T_{H}(t)$ that decreases monotonically
in time due to the energy dissipation in collisions. To describe fluctuations about the HCS, it is
convenient to introduce dimensionless length and time scales defined by
\begin{equation}
\label{2.1}
{\bm \ell} \equiv \frac{\bm r}{\lambda}
\end{equation}
and
\begin{equation}
\label{2.2}
s \equiv \int_{0}^{t} dt_{1}\, \frac{v_{0}(t_{1})}{\lambda},
\end{equation}
respectively. In the above expressions, $\lambda \equiv (n \sigma^{d-1})^{-1}$ and
\begin{equation}
\label{2.3}
v_{0}(t) = \left[ \frac{ 2 T_{H}(t)}{m} \right]^{1/2}.
\end{equation}
Then $\lambda$ is proportional to the mean free path of the gas in the HCS and $v_{0}(t)$ is a
characteristic thermal velocity. The granular temperature is defined from the average kinetic
energy of the grains with the Boltzmann constant formally set equal to unity.
It is easily verified that the time scale $s$ is proportional to the
accumulated average number of collisions per particle occurring in the system in the time interval
between $0$ and $t$. Consistently with Eqs. (\ref{2.1}) and (\ref{2.2}) a new velocity scale is introduced
through
\begin{equation}
\label{2.4}
{\bm c} \equiv \frac{\bm v}{v_{0}(t)}.
\end{equation}
Also, the deviation of $F({\bm r},{\bm v},t)$ from its average in the HCS, $f_{H}({\bm v},t)$,
\begin{equation}
\label{2.5}
\delta  F({\bm r},{\bm v},t) \equiv F({\bm r},{\bm v},t)-f_{H}({\bm v},t),
\end{equation}
is transformed to a dimensionless form by defining
\begin{equation}
\label{2.6}
\delta  \widetilde{F}({\bm \ell},{\bm c},s) \equiv n^{-1} v_{0}^{d}(t) \delta F({\bm r},{\bm v},t).
\end{equation}
This function obeys the Boltzmann-Langevin equation \cite{BMyG09}
\begin{equation}
\label{2.7}
\left[ \frac{\partial}{\partial s} +{\bm c} \cdot \frac{\partial}{\partial {\bm \ell}}- \Lambda ({\bm c}) \right]
\delta  \widetilde{F}({\bm \ell},{\bm c},s) = \widetilde{S} ( {\bm \ell},{\bm c},s),
\end{equation}
where $\Lambda ({\bm c})$ is the linearized inelastic Boltzmann operator given in Appendix \ref{ApA} and the
term $\widetilde{S} ({\bm \ell}, {\bm c},s)$ has the properties of a white noise term,
\begin{equation}
\label{2.8}
\langle \widetilde{S} ({\bm \ell}, {\bm c},s)\rangle_{H}=0,
\end{equation}
\begin{equation}
\label{2.9}
\langle \widetilde{S} ({\bm \ell}, {\bm c},s) \delta \widetilde{F}  ({\bm \ell}^{\prime}, {\bm c}^{\prime},s^{\prime})\rangle_{H} =0,
\end{equation}
for $s>s^{\prime}$, and
\begin{equation}
\label{2.10}
 \langle\widetilde{S} ({\bm \ell}, {\bm c},s)\widetilde{S} ({\bm \ell}^{\prime}, {\bm c}^{\prime},s^{\prime}) \rangle_{H} = n^{-1}
\lambda^{-d} \delta ( {\bm \ell} - {\bm \ell}^{\prime}) \delta (s-s ^{\prime}) \widetilde{\Gamma} ({\bm c}, {\bm c}^{\prime}).
\end{equation}
Here and in the following, angular brackets with the subindex $H$ are used to denote stochastic average in the HCS. The amplitude $\widetilde{\Gamma} ({\bm c}, {\bm c}^{\prime})$ of the noise is given by
\begin{equation}
\label{2.11}
\widetilde{\Gamma} ({\bm c},{\bm c}^{\prime}) =- \left[ \Lambda ({\bm c})+ \Lambda ({\bm c}^{\prime}) \right] \chi (c) \delta ({\bm c}
-{\bm c}^{\prime}) + \overline{T}_{0} ({\bm c},{\bm c}^{\prime}) \chi(c)  \chi(c^{\prime}),
\end{equation}
with $\overline{T}_{0} ({\bm c},{\bm c}^{\prime})$ being the inelastic binary collision operator given in Eq.\ (\ref{apa.1}) and $\chi(c)$ the dimensionless scaled velocity distribution of the HCS defined as
\begin{equation}
\label{2.12}
\chi(c) \equiv n^{-1} v_{0}^{d}(t) f_{H}({\bm v},t).
\end{equation}
An accurate expression for this distribution is known and it is reminded in Appendix \ref{ApA}, Eq. (\ref{apa.6}).
For the present purposes, it is useful to employ the Fourier representation and to introduce
\begin{equation}
\label{2.13}
\delta  \widetilde{F}({\bm k},{\bm c},s)  \equiv \int d{\bm \ell}\, e^{-i {\bm k} \cdot {\bm \ell}} \delta  \widetilde{F}({\bm \ell},{\bm c},s).
\end{equation}
In Fourier space, Eq.\ (\ref{2.7}) becomes
\begin{equation}
\label{2.14}
\left[ \frac{\partial}{\partial s} - \Lambda ({\bm k},{\bm c} ) \right] \delta  \widetilde{F}({\bm k},{\bm c},s)=
\widetilde{S} ({\bm k}, {\bm c},s),
\end{equation}
where
\begin{equation}
\label{2.15}
\Lambda ({\bm k},{\bm c} )  \equiv \Lambda ({\bm c} ) - i {\bm k} \cdot {\bm c}
\end{equation}
is the linear inhomogeneous inelastic Boltzmann operator. The property of the noise in Eq. (\ref{2.10}) now reads
\begin{equation}
\label{2.16}
\langle \widetilde{S} ({\bm k}, {\bm c},s) \widetilde{S} ({\bm k}^{\prime}, {\bm c}^{\prime},s^{\prime})\rangle_{H} =
\frac{\widetilde{V}^{2}}{N} \delta_{{\bm k},-{\bm k}^{\prime}}
 \delta (s-s ^{\prime}) \widetilde{\Gamma} ({\bm c}, {\bm c}^{\prime}),
\end{equation}
$\delta_{{\bm k},-{\bm k}^{\prime}} $ being the Kronecker delta symbol and $\widetilde{V} \equiv
V \lambda^{-d}$ the volume of the system in the length scale defined by $\lambda$.

\section{balance equations for the fluctuating hydrodynamic fields}
\label{s3}
Consider the mesoscopic number of particles density $N({\bm r},t)$, momentum density ${\bm G}({\bm r},t)$, and
energy density $E({\bm r},t)$. Dimensionless deviations from their average values in the HCS are introduced
through the definitions
\begin{equation}
\label{3.1} \delta \rho ({\bm k},s) \equiv \frac{N ({\bm
k},s)-n}{n} = \int d{\bm c}\, \delta \widetilde{F} ( {\bm k},{\bm
c},s),
\end{equation}
\begin{equation}
\label{3.2} \delta {\bm \omega} ({\bm k},s) \equiv \frac{\delta {\bm
G}({\bm k},s)}{m n v_{0}(t)} = \int d{\bm c}\, {\bm c} \delta
\widetilde{F} ( {\bm k},{\bm c},s),
\end{equation}
\begin{equation}
\label{3.3} \delta \epsilon ({\bm k},s) \equiv \frac{2}{d n T_{H}(t)}\, [E({\bm k},s)- \frac{d}{2} n T_{H}(s)]
 = \frac{2}{d} \int d{\bm c}\, c^{2} \delta
\widetilde{F} ( {\bm k},{\bm c},s).
\end{equation}
In \cite{BMyG09}, balance equations for these fields were derived by taking velocity moments in the
fluctuating Boltzmann equation. In Fourier space they read
 \begin{equation}
\label{3.4}
\frac{\partial}{\partial s}\, \delta \rho ({\bm k},s)+i {\bm k} \cdot \delta
{\bm \omega} ({\bm k},s)=0,
\end{equation}
\begin{equation}
\label{3.5} \left( \frac{\partial}{\partial s} -\frac{\zeta_{0}}{2}
\right) \delta {\bm \omega} ({\bm k},s) + i {\bm k} \cdot \delta {\sf \Pi} ({\bm k},s)=0,
\end{equation}
\begin{equation}
\label{3.6} \frac{\partial}{\partial s}\, \delta \epsilon ({\bm k},s) +  i\, \frac{d+2}{d}\,
{\bm k} \cdot \delta {\bm \omega} ({\bm
k},s) + \delta \zeta({\bm k},s)  +  i\, \frac{2}{d}
{\bm k} \cdot \delta {\bm \phi} ({\bm
k},s) = \widetilde{S}_{\epsilon}({\bm k},s).
\end{equation}
In the above equations, $\zeta_{0}$ is the cooling rate of the HCS. Its definition is given in
Appendix \ref{ApA}, where an approximated expression is also provided. Moreover,
$\delta {\sf \Pi}({\bm k}, s)$ and $\delta
{\bm \phi}({\bm k},s)$ are the fluctuating pressure tensor and heat
flux, respectively. They are functionals of the fluctuating distribution function,
\begin{equation}
\label{3.7} \delta {\sf \Pi}({\bm k}, s) \equiv \frac{\delta \epsilon
({\bm k},s)}{2} {\sf I}+ \int d{\bm c}\, {\sf \Delta} ({\bm c})
\delta \widetilde{F} ( {\bm k},{\bm c},s),
\end{equation}
\begin{equation}
\label{3.8} \delta {\bm \phi}({\bm k},s) = \int d{\bm k}\, {\bm
\Sigma} ({\bm c}) \delta \widetilde{F} ( {\bm k},{\bm c},s),
\end{equation}
where ${\sf I}$ is the unit tensor of dimension $d$ and
\begin{equation}
\label{3.9} {\sf \Delta}({\bm c}) \equiv {\bm c} {\bm c}-
\frac{c^{2}}{d}\ {\sf I},
\end{equation}
\begin{equation}
\label{3.10} {\bm \Sigma}({\bm c}) \equiv \left( c^{2}-
\frac{d+2}{2} \right) {\bm c}.
\end{equation}
The equation for the fluctuating energy field, Eq.\ (\ref{3.6}), involves two terms
vanishing in the elastic limit $\alpha \rightarrow 1$. One is associate with the cooling rate
fluctuations
\begin{equation}
\label{3.11}
\delta \zeta ({\bm k},s) = - \frac{2}{d} \int d{\bm c}\, c^{2} \Lambda ({\bm c}) \delta
\widetilde{F} ({\bm k},{\bm c},s),
\end{equation}
and the other one is the noise term   $\widetilde{S}_{\epsilon}({\bm k},s)$ arising directly from the
fluctuations in phase space,
\begin{equation}
\label{3.12}
\widetilde{S}_{\epsilon}({\bm k},s) = \frac{2}{d} \int d{\bm c}\, c^{2} \widetilde{S} ({\bm k},{\bm c},s).
\end{equation}
From Eqs. (\ref{2.8}) and (\ref{2.16}) it follows that
\begin{equation}
\label{3.13}
\langle \widetilde{S}_{\epsilon}({\bm k},s)
\rangle_{\text{H}} =0
\end{equation}
and
\begin{equation}
\label{3.14} \langle  \widetilde{S}_{\epsilon}({\bm k},s)
\widetilde{S}_{\epsilon}({\bm
k}^{\prime},s^{\prime})\rangle_{\text{H}} = \frac{4 \widetilde{V}^{2}}{d^{2} N} \delta_{{\bm k},-{\bm k}^{\prime}}
\delta( s- s^{\prime}) \int d{\bm c} \int d {\bm c}^{\prime} c^{2} c^{\prime 2} \widetilde{\Gamma} ({\bm c}, {\bm c}^{\prime}).
\end{equation}

Also it is
\begin{equation}
\label{3.15}
\langle \widetilde{S}_{\epsilon}({\bm k},s) \widetilde{F} ({\bm k}^{\prime},{\bm c}^{\prime},s^{\prime})\rangle_{H}=0
\end{equation}
and
\begin{equation}
\label{3.16}
\langle \widetilde{S}_{\epsilon}({\bm k},s) \delta \rho ({\bm k}^{\prime},s^{\prime})\rangle_{H}
= \langle \widetilde{S}_{\epsilon}({\bm k},s) \delta {\bm \omega} ({\bm k}^{\prime},s^{\prime})\rangle_{H}
= \langle \widetilde{S}_{\epsilon}({\bm k},s) \delta \epsilon ({\bm k}^{\prime},s^{\prime})\rangle_{H}
=0,
\end{equation}
for $s> s^{\prime}$.

Equations (\ref{3.4})-(\ref{3.6}) are not closed, since they contain the quantities $\delta {\bm \Pi}$, $\delta
{\bm \phi}$, and $\delta \zeta$, defined above in terms of $\delta \widetilde{F} ( {\bm k},{\bm c},s)$, as well
as the noise term $\widetilde{S}_{\epsilon} ({\bm k},s)$. The aim in the following will be to obtain a self-consistent
description for the fluctuating fields. In this context, it will be useful to consider the eigenvalue problem for the inelastic
homogeneous linear Boltzmann operator \cite{BDyR03,ByD05},
\begin{equation}
\label{3.17}
\Lambda ({\bm c}) \xi_{\beta} ({\bm c})= \lambda_{\beta} \xi_{\beta} ({\bm c}).
\end{equation}
The solutions of this equation corresponding to the infinite wave length limit ($k=0$) of the hydrodynamic
equations are given by \cite{BDyR03,ByD05}
\begin{equation}
\label{3.18}
\lambda_{1}=0, \quad \lambda_{2}=\frac{\zeta_{0}}{2}\, ,
\quad \lambda_{3} =-\frac{\zeta_{0}}{2},
\end{equation}
\begin{equation}
\label{3.19} \xi_{1}({\bm c})= \chi(c)+\frac{\partial}{\partial {\bm
c}} \cdot \left[ {\bm c} \chi (c) \right], \quad {\bm \xi}_{2}({\bm
c})=-\frac{\partial \chi(c)}{\partial {\bm c}}, \quad \xi_{3}({\bm
c})= -\frac{\partial}{\partial {\bm c}} \cdot \left[ {\bm c} \chi
(c) \right].
\end{equation}
The eigenvalue $\lambda_{2}$ is $d$-fold degenerate. Normalization of the distribution
function requires that the velocity distribution functions being considered be integrable.
However, the strongest requirement is made now that they be elements of a Hilbert space
with scalar product defined as
\begin{equation}
\label{3.20}
\langle g|h\rangle \equiv \int d{\bm c}\, \chi^{-1} (c)
g^{*}({\bm c}) h({\bm c}),
\end{equation}
where $g^{*}({\bm c})$ is the complex conjugate of $g({\bm c})$. The operator $\Lambda
({\bm c})$ is not Hermitian and the eigenfunctions $\xi_{\beta}({\bm c})$ are not orthogonal.
Then, it is convenient to introduce a set of functions $\overline{\xi}_{\beta} ({\bm c})$ biorthogonal
to the above eigenfunctions, i.e. verifying
\begin{equation}
\label{3.21}
\langle \overline{\xi}_{\beta} |\xi_{\beta^{\prime}}
\rangle =\delta_{\beta,\beta^{\prime}}\, .
\end{equation}
A convenient choice is \cite{BDyR03,ByD05}
\begin{equation}
\label{3.21a} \overline{\xi}_{1}({\bm c})=\chi(c), \quad \overline{\bm
\xi}_{2} ({\bm c})={\bm c} \chi(c), \quad \overline{\xi}_{3} ({\bm
c})= \left( \frac{c^{2}}{d} +\frac{1}{2} \right) \chi(c).
\end{equation}
Using the bi-orthogonal sets of functions, a projection operator $\mathcal{P}$ over the hydrodynamic part of the Hilbert space can
be defined by
\begin{equation}
\label{3.22}
 \mathcal{P} g({\bm c}) \equiv \sum_{\beta=1}^{d+2}
\xi_{\beta}({\bm c}) \langle\overline{\xi}_{\beta}|g\rangle.
\end{equation}
By means of $\mathcal{P}$, the fluctuating one-particle distribution function can be decomposed
into its hydrodynamic and non-hydrodynamic components,
\begin{equation}
\label{3.23}
\delta \widetilde{F} ({\bm k},{\bm c},s)= \mathcal{P}
\delta \widetilde{F} ({\bm k},{\bm c},s) +\mathcal{P}_{\perp}
\delta \widetilde{F} ({\bm k},{\bm c},s),
\end{equation}
where $\mathcal{P}_{\perp} \equiv 1- \mathcal{P}$. Application of $\mathcal{P}_{\perp} $ to
both sides of Eq. (\ref{2.14}) and formal integration of the resulting equation yields
\cite{BMyG09}
\begin{eqnarray}
\label{3.24} \mathcal{P}_{\perp} \delta \widetilde{F}({\bm
k},{\bm c},s) & = & \mathcal{U}({\bm k},{\bm c},s)
\mathcal{P}_{\perp}
\delta \widetilde{F} ({\bm k},{\bm c},0)+ \int_{0}^{s} ds^{\prime}\, \mathcal{U}({\bm k},{\bm
c},s^{\prime})  \mathcal{P}_{\perp} \left[ - i {\bm k} \cdot {\bm c}
\mathcal{P} \delta \widetilde{F} ({\bm k},{\bm c},s-s^{\prime})
\right. \nonumber \\
&& \left.+ \widetilde{S}({\bm k},{\bm c},s-s^{\prime}) \right],
\end{eqnarray}
where
\begin{equation}
\label{3.25} \mathcal{U}({\bm k},{\bm c},s) \equiv \exp \left[s
\mathcal{P}_{\perp}  \Lambda ({\bm k},{\bm c}) \mathcal{P}_{\perp} \right].
\end{equation}

The hypothesis is made now that for large enough $s$, the first term on the right
hand side of Eq.\ (\ref{3.24}) becomes negligible. This is related with the ageing to
hydrodynamics, implying that the non-hydrodynamic part of the initial condition is forgotten
on the hydrodynamic time scale. In addition, the limit of small wavevector ${\bm k}$ is considered
and only terms up to first order in it are kept. Then, for large values of $s$, Eq.\ (\ref{3.24}) becomes
\begin{eqnarray}
\label{3.26} \mathcal{P}_{\perp} \delta \widetilde{F} ({\bm
k},{\bm c},s) & \simeq & \int_{0}^{s} ds^{\prime}\, \mathcal{P}_{\perp}
e^{ s^{\prime}\Lambda ({\bm c})} \left( - i {\bm k} \cdot {\bm c}
\right)  \mathcal{P} \delta
\widetilde{F} ({\bm k},{\bm c},s- s^{\prime}) \nonumber \\
& & + \int_{0}^{s} d s^{\prime}\,  \mathcal{U}({\bm k},{\bm
c},s^{\prime}) \mathcal{P}_{\perp} \widetilde{S}({\bm k},{\bm
c},s-s^{\prime}), \nonumber \\
\end{eqnarray}
This expression will be used in the next sections to obtain explicit expression for
the fluctuating fluxes and for the cooling rate in the Navier-Stokes approximation.

\section{Langevin equation for the velocity field}
\label{s4}
By direct calculation, it is easily verified that
\begin{equation}
\label{4.1}
\int d{\bm c}\, {\sf \Delta}  ({\bm c}) \xi_{\beta}({\bm c})=0.
\end{equation}
Here $\xi_{\beta}({\bm c})$ stands for any of the hydrodynamic eigenfunctions of $\Lambda
({\bm c})$ given in Eq. (\ref{3.19}).  Therefore,
\begin{equation}
\label{4.2}
\int d{\bm c}\, \Delta ({\bm c}) \mathcal{P} \delta \widetilde{F}
({\bm k},{\bm c},s)=0
\end{equation}
and, consequently, $\delta \widetilde{F}$ can be substituted by $ \mathcal{P}_{\perp}
\delta \widetilde{F}$ on the right hand side of the expression for the fluctuating pressure tensor,
Eq.\ (\ref{3.7}). Using next Eq.\
(\ref{3.26}) gives
\begin{equation}
\label{4.3}
\delta {\sf \Pi} ({\bm k},s) \simeq \frac{ \delta \epsilon ({\bm k},s)}{2}\, \sf{I} + \delta_{1}
{\sf \Pi} ({\bm k},s) + {\sf R} ({\bm k},s),
\end{equation}
where
\begin{equation}
\label{4.3a} \delta_{1} {\sf \Pi} ({\bm k},s) =   \int_{0}^{s} d
s^{\prime}\, \int d{\bm c}\, {\sf \Delta}({\bm c}) e^{s^{\prime}
\Lambda ({\bm c})} (-i {\bm k} \cdot {\bm c} )   \mathcal{P} \delta
\widetilde{F} ({\bm k}, {\bm c},s-
s^{\prime}),
\end{equation}
and
\begin{equation}
\label{4.4} {\sf R}({\bm k},s)= \int_{0}^{s} ds^{\prime} \int d{\bm
c}\, {\sf \Delta} ({\bm c}) \mathcal{U} ({\bm k},{\bm c},
s^{\prime}) \mathcal{P}_{\perp} \widetilde{S} ({\bm k},{\bm
c},s-s^{\prime}).
\end{equation}
Equation  (\ref{3.5}) implies that
\begin{equation}
\label{4.5}
\delta {\bm \omega} ({\bm k},s-s^{\prime}) \simeq  e^{-s^{\prime}
\zeta_{0}/2}\delta {\bm \omega} ({\bm k},s)
\end{equation}
to lowest order in ${\bm k}$. Using this, it is obtained that \cite{BMyG09}
\begin{equation}
\label{4.6}
\delta_{1} {\Pi}_{ij} ({\bm k},s) = -i
\widetilde{\eta}(s) \left[ k_{i} \delta \omega_{j} ({\bm k},s) +
k_{j} \delta \omega_{i} ({\bm k},s) - \frac{2}{d}\, \delta_{ij} {\bm k} \cdot \delta
{\bm \omega} ({\bm k},s) \right],
\end{equation}
 valid to first order in $k$. In the above expression, $\widetilde{\eta} (s)$ is  the
 time-dependent dimensionless shear viscosity of the HCS \cite{DyB02},
 \begin{equation}
 \label{4.7}
 \widetilde{\eta}(s) \equiv \frac{1}{d^2+d-2} \sum_{i,j}^{d} \int d{\bm c}\,
 \Delta_{ij}({\bm c}) \Phi_{2,ij} ({\bm c},s),
 \end{equation}
 \begin{equation}
 \label{4.8}
\Phi_{2,ij} ({\bm c},s)= \int_{0}^{s} ds^{\prime}\, e^{s^{\prime} \left( \Lambda -\zeta_{0}/2
\right)} \xi_{2,i}({\bm c}) c_{j}.
\end{equation}
Equation (\ref{4.7}) agrees with the result found from the nonlinear Boltzmann equation by using the
Chapman-Enskog algorithm, and its long time limit has been evaluated in the first Sonine approximation
\cite{BDKyS98,ByC01}.

The last term on the right hand side of Eq. (\ref{4.3}) is defined in Eq. (\ref{4.4}) and it has the property
\begin{equation}
\label{4.9}
\langle {\sf R} ({\bm k},s)\rangle_{H}=0,
\end{equation}
that follows from Eq. (\ref{2.8}). Moreover, in Appendix \ref{ApB} it is shown that for $s\gg 1$, $s^{\prime} \gg 1$ and to lowest order in $k$, it is
\begin{equation}
\label{4.10}
 \langle R_{ij}({\bm k},s)R_{lm}({\bm k}^{\prime},s^{\prime})\rangle_{H} = \frac{\widetilde{V}^{2}}{N} \delta_{{\bm k},-{\bm k}^{\prime}} G(|s-s^{\prime}|) ( \delta_{il} \delta_{jm}+ \delta_{im} \delta_{jl}
- \frac{2}{d} \delta_{ij} \delta_{lm} ),
\end{equation}
where
\begin{equation}
\label{4.11}
G(s) \equiv \frac{1}{d^{2}+d-2} \sum_{i,j}^{d} \int d{\bm c}_{1} \int d{\bm c}_{2}\, \Delta_{ij} ({\bm c}_{1})
\Delta_{ij}({\bm c}_{2}) e^{s \Lambda ({\bm c}_{2})} \widetilde{\phi}_{H}({\bm c}_{1},{\bm c}_{2}),
\end{equation}
with $\widetilde{\phi}_{H}({\bm c}_{1},{\bm c}_{2})$ being the solution of the equation
\begin{equation}
\label{4.12}
\left[ \Lambda ({\bm c}_{1}) + \Lambda ({\bm c}_{2}) \right] \widetilde{\phi}_{H}({\bm c}_{1},{\bm c}_{2}) =
- \mathcal{P}_{\perp}^{(1)} \mathcal{P}_{\perp}^{(2)} \widetilde{\Gamma} ({\bm c}_{1}, {\bm c}_{2} ).
\end{equation}
The operators $\mathcal{P}^{(1)}_{\perp}$ and $\mathcal{P}_{\perp}^{(2)}$ are defined like $\mathcal{P}_{\perp}$, but acting on functions of the velocities ${\bm c}_{1}$ and ${\bm c}_{2}$, respectively. An approximation will be introduced at this point. The noise amplitude $\widetilde{\Gamma} ({\bm c}_{1},{\bm c}_{2})$ given by Eq.\ (\ref{2.11}) has two contributions of a rather different physical origin. The first one reflects fluctuations induced by collisions as a consequence of two different particles colliding independently at the same position with the same environment. This contribution does not vanishes even for a molecular gas at equilibrium \cite{ByZ69}.
On the other hand, the second contribution to $\widetilde{\Gamma} ({\bm c}_{1},{\bm c}_{2})$ is directly related with the velocity correlations between particles in the HCS and vanishes for an ordinary fluid at equilibrium. Here, the hypothesis will be made that only the hydrodynamic component of the velocity correlations, as extracted by the operator $ \mathcal{P}^{(1)}\mathcal{P}^{(2)}$, is relevant, while the remaining kinetic or non-hydrodynamic  component is negligible. Some justifications for this assumption are provided in the discussion section of the paper.  Then, the second term on the right hand side of Eq. (\ref{2.11}) is neglected when substituting it into Eq.\ (\ref{4.12}) and the solution of this equation can be written down by simple inspection,
\begin{equation}
\label{4.13}
\widetilde{\phi}_{H} ({\bm c}_{1},{\bm c}_{2}) \simeq \mathcal{P}^{(1)}_{\perp} \mathcal{P}^{(2)}_{\perp} \chi (c_{1}) \delta
({\bm c}_{1}-{\bm c}_{2}).
\end{equation}
Use of Eqs. (\ref{4.3}) and (\ref{4.6}) into Eq. (\ref{3.5}) gives the fluctuating Navier-Stokes equation for the
velocity field,
\begin{equation}
\label{4.14}
\left( \frac{\partial}{\partial s} - \frac{\zeta_{0}}{2} \right) \delta {\bm \omega} ({\bm k},s) + i {\bm k}
\frac{\delta \epsilon ({\bm k},s)}{2} + \widetilde{\eta} k^{2} \left[ \delta {\bm \omega} ({\bm k},s) + \frac{d-2}{d} \widehat{\bm k} \cdot \delta {\bm \omega}({\bm k},s) \widehat{\bm k} \right] = \widetilde{\bm W} ({\bm k},s),
\end{equation}
where $\widehat{\bm k} \equiv {\bm k}/k$ and the noise term $\widetilde{\bm W}({\bm k},s) = -i {\bm k} \cdot {\sf R} ({\bm k},s)$ has the properties
\begin{equation}
\label{4.15}
\langle \widetilde{\bm W} ({\bm k},s)\rangle_{H} =0,
\end{equation}
\begin{equation}
\label{4.16}
\langle \widetilde{W}_{i} ({\bm k},s) \widetilde{W}_{j} ({\bm k}^{\prime} ,s^{\prime}) \rangle_{H} =  \frac{\widetilde{V}^{2}}{N}
\delta_{{\bm k}, -{\bm k}^{\prime}} G(|s-s^{\prime}|) k^{2} \left( \delta_{ij}+ \frac{d-2}{d} \widehat{k}_{i} \widehat{k}_{j} \right).
\end{equation}
The quantity $G(s)$ is evaluated approximately in Appendix \ref{ApC}. The used approximation, which is an exact property for the Maxwell model for inelastic gases \cite{BGyM10} reads
\begin{equation}
\label{4.17}
\Lambda^{+} ({\bm c}) \Delta_{xy}({\bm c}) \chi (c) \simeq \overline{\lambda}_{4} \Delta_{xy} ({\bm c}) \chi (c).
\end{equation}
In the above relation, $\Lambda^{+}$ is the adjoint operator of $\Lambda$ and the eigenvalues $\overline{\lambda}_{4}$ is determined self-consistently. The result is
\begin{equation}
\label{4.18}
G(s) \simeq \frac{1+a_{2}(\alpha)}{4} e^{s \overline{\lambda}_{4}},
\end{equation}
with
\begin{equation}
\label{4.19}
\overline{\lambda}_{4} = \zeta_{0} + \frac{4I(\alpha)}{1+a_{2}(\alpha)}.
\end{equation}
The explicit form of the functions $a_{2}(\alpha)$ and $I(\alpha)$ is given in Eqs. (\ref{apa.8}) and (\ref{apc.6}), respectively.

It is interesting to evaluate the shear viscosity $\widetilde{\eta}$ given in Eq.\ (\ref{4.7}) in the same approximation as employed to calculate the function $G(s)$, i.e. by using Eq.\ (\ref{4.17}). In the long time limit
$s \gg 1$ it is found in Appendix \ref{ApC} that
\begin{equation}
\label{4.20}
\widetilde{\eta} \simeq \left( \frac{8 |I(\alpha)|}{1+ a_{2} (\alpha)}- \zeta_{0} \right)^{-1}.
\end{equation}
This result is very close to the one reported in refs. \cite{BDKyS98} and \cite{ByC01}, both being practically indistinguishable for $\alpha \agt 0.6$. Moreover, Eq. (\ref{4.20}) coincides with the expression derived
in \cite{GSyM07} by using a modified Sonine expansion in which the gaussian is replaced by the (approximated) distribution of the HCS, $\chi (c)$.

\section{Langevin equation for the energy field}
\label{s5}
First consider the term $\delta \zeta ({\bm k},s)$ given in Eq.\ (\ref{3.11}). It will be evaluated here by making
an approximation in the same spirit as Eq. (\ref{4.17}), namely by writing
\begin{equation}
\label{5.1}
\Lambda^{+}({\bm c}) \overline{\xi}_{3}({\bm c}) \simeq \lambda_{3} \overline{\xi}_{3} ({\bm c}) = - \frac{\zeta_{0}}{2}\, \overline{\xi}_{3}({\bm c}).
\end{equation}
Then, by realizing that Eq.\ (\ref{3.11}) is equivalent to
\begin{equation}
\label{5.2}
\delta \zeta ({\bm k},s) = - 2 \langle \overline{\xi}_{3}|\Lambda \delta \widetilde{F}\rangle,
\end{equation}
it follows that
\begin{equation}
\label{5.3}
\delta \zeta ({\bm k},s) \simeq \frac{\zeta_{0}}{2} \left[ \delta \epsilon ({\bm k},s)+ \delta \rho ({\bm k},s) \right].
\end{equation}
The same result is obtained if, instead of Eq.\ (\ref{5.1}), the approximation
\begin{equation}
\label{5.4}
 \langle\overline{\xi}_{3}|\Lambda \delta \widetilde{F}\rangle \simeq \langle \overline{\xi}_{3}|\Lambda \mathcal{P}\delta \widetilde{F}\rangle
\end{equation}
is employed. Therefore contributions to $\delta \zeta ({\bm k},s)$ from $\mathcal{P}_{\perp} \delta \widetilde{ F}$ are being neglected in Eq.\ (\ref{5.3}). This includes, in particular, terms proportional to the gradients of the fluctuating hydrodynamic fields (see, for instance, Eq.\ (\ref{3.24})). As for Eq.\ (\ref{4.17}), the approximation given by Eq. (\ref{5.1}) becomes an exact relation for the inelastic Maxwell model for granular gases \cite{BGyM10}. Moreover, it has been shown that the linear transport coefficients associated with the gradient expansion of the average cooling rate are very small as compared with the similar contributions coming from the hydrodynamic fluxes \cite{BDKyS98}. Actually, the only contribution to the hydrodynamic equations from the average cooling rate that is kept in practically all the literature is the one of zeroth order in the gradients.

Next, the heat flux term $\delta {\bm \phi}$ defined in Eq.\ (\ref{3.8}) has to be evaluated. It contains the function ${\bf \Sigma}({\bm c})$ that verifies a relationship similar to Eq.\ (\ref{4.1}),
\begin{equation}
\label{5.5}
\int d{\bm c}\, {\bm \Sigma}({\bm c}) \xi_{\beta} ({\bm c}) =0,
\end{equation}
for all the hydrodynamic modes $\xi_{\beta}({\bm c})$. Therefore, Eq. (\ref{3.8}) is equivalent to
\begin{equation}
\label{5.6}
\delta {\bm \phi} ({\bm k},s) = \int d{\bm c}\, {\bm \Sigma}({\bm c}) \mathcal{P}_{\perp} \delta \widetilde{F} ({\bm k},{\bm c},s),
\end{equation}
and by means  of Eq.\ (\ref{3.26}), the energy flux can be decomposed as
\begin{equation}
\label{5.7}
\delta {\bm \phi} ({\bm k},s) = \delta_{1} {\bm \phi} ({\bm k},s) + {\bm Z} ({\bm k},s),
\end{equation}
where
\begin{equation}
\label{5.8}
\delta_{1} {\bm \phi} ({\bm k},s)= \int_{0}^{s} ds^{\prime} \int d{\bm c}\,  {\bm \Sigma} ({\bm c}) e^{s^{\prime}
\Lambda ({\bm c})} (-i {\bm k} \cdot {\bm c}) \mathcal{P} \delta \widetilde{F} ({\bm k}, {\bm c}, s- s^{\prime}),
\end{equation}
and
\begin{equation}
\label{5.9}
{\bm Z} ({\bm k},s)= \int_{0}^{s} ds^{\prime} \int d{\bm c}\, {\bm \Sigma} ({\bm c}) \mathcal{U}({\bm k},{\bm c},s^{\prime})
\mathcal{P}_{\perp} \widetilde{S} ({\bm k},{\bm c}, s-s^{\prime}).
\end{equation}
The above expressions are valid up to first order in $k$. A direct calculation gives
\begin{equation}
\label{5.10}
\mathcal{P} \delta \widetilde{F} ({\bm k}, {\bm c}, s)= \delta \rho ({\bm k},s) \left[ \xi_{1} ({\bm c})+ \frac{\xi_{3}({\bm c})}{2} \right] + \delta {\bm \omega} ({\bm k},s) \cdot {\bm \xi}_{2} ({\bm c})+ \frac{1}{2}\, \delta \epsilon ({\bm k},s) \xi_{3} ({\bm c}),
\end{equation}
and taking into account the isotropy of the operator $\Lambda ({\bm c})$, Eq. (\ref{5.8}) can be rewritten as
\begin{eqnarray}
\label{5.11}
\delta_{1} {\bm \phi} ({\bm k},s) &= &\int_{0}^{s} ds^{\prime} \int d{\bm c}\,  {\bm \Sigma} ({\bm c}) e^{s^{\prime}
\Lambda ({\bm c})} (-i {\bm k} \cdot {\bm c}) \nonumber \\
&& \times \left\{ \delta \rho ({\bm k},s-s^{\prime}) \xi_{1} ({\bm c})+
\frac{1}{2} \left[ \delta \epsilon ({\bm k},s-s^{\prime})+ \delta \rho ({\bm k},s-s^{\prime}) \right] \xi_{3} ({\bm c}) \right\}.
\end{eqnarray}
The next task is to evaluate the functions $\delta \rho ({\bm k},s-s^{\prime})$ and $\delta \epsilon ({\bm k},s-s^{\prime})$ as functions of the fluctuating hydrodynamic fields at time $s$ to lowest (zeroth) order in $k$, in order to have an expression for $\delta_{1} {\bm \phi} ({\bm k},s)$ valid to first order in $k$. Use of the balance hydrodynamic equations (\ref{3.4})-(\ref{3.6}) leads to
\begin{equation}
\label{5.12}
\delta \rho ({\bm k},s-s^{\prime}) \simeq \delta \rho ({\bm k},s),
\end{equation}
\begin{eqnarray}
\label{5.13}
\delta \epsilon ({\bm k},s-s^{\prime})+ \delta \rho ({\bm k},s-s^{\prime}) & \simeq & e^{\zeta_{0} s^{\prime}/2}
\left[ \delta \epsilon ({\bm k},s)+ \delta \rho ({\bm k},s) \right] \nonumber \\
& & - e^{- \zeta_{0} (s- s^{\prime})/2} \int_{s-s^{\prime}}^{s} ds_{1}\ e^{\zeta_{0}s _{1}/2} \widetilde{S}_{\epsilon}({\bm k},s_{1}).
\end{eqnarray}
When the above expressions are substituted into Eq.\ (\ref{5.11}), two contributions physically rather different are identified. One of them is of hydrodynamic character and similar to the expression for the fluctuating heat flux for molecular gases derived by Landau and Lifshitz \cite{LyL66}, while the other one is an intrinsic noise term following directly from the inelasticity of collisions,
\begin{equation}
\label{5.14}
\delta_{1} {\bm \phi} ({\bm k}, s) = \delta_{1}^{(H)} {\bm \phi} ({\bm k}, s)+ \delta_{1}^{(I)} {\bm \phi} ({\bm k},s).
\end{equation}
After simple manipulations, the first contribution can be expressed as
\begin{eqnarray}
\label{5.15}
\delta_{1}^{(H)} {\bm \phi} ({\bm k}, s) &= &\int_{0}^{s} ds^{\prime} \int d{\bm c}\,  {\bm \Sigma} ({\bm c}) e^{s^{\prime}
\Lambda ({\bm c})} (-i {\bm k} \cdot {\bm c}) \nonumber \\
&& \times \left\{ \delta \rho ({\bm k},s) \xi_{1} ({\bm c})+
\frac{1}{2} e^{\zeta_{0}s^{\prime}/2}
\left[ \delta \rho ({\bm k},s)+ \delta \epsilon ({\bm k},s) \right] \xi_{3} ({\bm c}) \right\}
\nonumber \\
&=& -\widetilde{\kappa} i {\bm k} \delta \epsilon ({\bm k},s) - \left( \widetilde{\mu} - \widetilde{\kappa} \right) i {\bm k} \delta \rho ({\bm k},s).
\end{eqnarray}
The right hand side of this equation  has the same form as the (generalized) Fourier law for dilute granular gases \cite{BDKyS98,ByC01}. It involves two transport coefficients: the (thermal) heat conductivity $\widetilde{\kappa}$ and the diffusive heat conductivity $\widetilde{\mu}$. Their expressions are:
\begin{equation}
\label{5.16}
\widetilde{\kappa}(s) = \frac{1}{d} \int d{\bm c}\,  {\bm \Sigma} ({\bm c}) \cdot {\bm \Phi}_{3} ({\bm c},s),
\end{equation}
\begin{equation}
\label{5.17}
\widetilde{\mu}(s) = \frac{1}{d} \int d{\bm c}\,  {\bm \Sigma} ({\bm c}) \cdot {\bm \Phi}_{1} ({\bm c},s),
\end{equation}
with
\begin{equation}
\label{5.18}
{\bm \Phi}_{1} ({\bm c},s) = \int_{0}^{s} ds^{\prime}\, e^{s^{\prime} \Lambda ({\bm c})} \xi_{1} ({\bm c}) {\bm c}
+2 {\bm \Phi}_{3} ({\bm c},s),
\end{equation}
\begin{equation}
\label{5.19}
{\bm \Phi}_{3} ({\bm c},s)= \frac{1}{2} \int_{0}^{s} ds^{\prime}\, e^{s^{\prime} \left[ \Lambda ({\bm c}) + \zeta_{0}/2 \right]} {\xi}_{3} ({\bm c}) {\bm c}.
\end{equation}
As usual, $\widetilde{\kappa}(s)$ and $\widetilde{\mu}(s)$ are expected to reach steady plateau values for large enough $s$, when the hydrodynamic description is accurate. Both transport coefficients have been evaluated in the first Sonine approximation \cite{BDKyS98,ByC01}.

The noise term in Eq. (\ref{5.14}) is given by
\begin{eqnarray}
\label{5.20}
\delta_{1}^{(I)} {\bm \phi} ({\bm k},s)& = &\frac{1}{2} \int_{0}^{s} ds^{\prime} \int d{\bm c}\,  {\bm \Sigma} ({\bm c}) e^{s^{\prime}
\Lambda ({\bm c})} i {\bm k} \cdot {\bm c} e^{-\zeta_{0}(s-s^{\prime})/2} \nonumber \\
&& \times \int_{s-s^{\prime}}^{s} ds_{1}\ e^{\zeta_{0}s _{1}/2} \widetilde{S}_{\epsilon}({\bm k},s_{1}) \xi_{3} ({\bm c}).
\end{eqnarray}
Substitution of Eqs.\ (\ref{5.3}), (\ref{5.7}), (\ref{5.14}) and (\ref{5.15}) into Eq.\ (\ref{3.6}) yields
\begin{eqnarray}
\label{5.21}
\left( \frac{\partial}{\partial s} + \frac{\zeta_{0}}{2} \right) \delta \epsilon ({\bm k},s)+ \frac{\zeta_{0}}{2}
\delta \rho ({\bm k},s) + i\, \frac{d+2}{d} {\bm k}\cdot \delta {\bm \omega} ({\bm k},s)  \nonumber \\
+ \frac{2}{d}\, k^{2}
\left[ \widetilde{\kappa} \delta \epsilon ({\bm k},s) + ( \widetilde{\mu} - \widetilde{\kappa}) \delta \rho ({\bm k},s) \right] = \widetilde{\mathcal E}({\bm k},s),
\end{eqnarray}
where
\begin{equation}
\label{5.22}
\widetilde{\mathcal E}({\bm k},s) \equiv \widetilde{S}_{\epsilon} ({\bm k},s) - \frac{2i}{d} {\bm k} \cdot {\bm Z} ({\bm k},s)- \frac{2i}{d}\, {\bm k} \cdot \delta_{1}^{(I)} {\bm \phi} ({\bm k},s)
\end{equation}
is identified as the total noise term. The functions  ${\bm Z} ({\bm k},s)$ and $\delta_{1}^{(I)} {\bm \phi} ({\bm k},s)$ are defined in Eqs.\ (\ref{5.9}) and (\ref{5.20}), respectively, and the intrinsic inelastic noise term
$\widetilde{S}_{\epsilon} ({\bm k},s)$ is given in Eq.\ (\ref{3.12}). It is trivial to verify that
\begin{equation}
\label{5.23}
\langle \widetilde{\mathcal E}({\bm k},s)\rangle_{H}=0,
\end{equation}
while the corresponding correlation function  is calculated in Appendices \ref{ApD} and \ref{ApE} under well defined and controlled approximations, which will be made explicit also below. The result reads
\begin{eqnarray}
\label{5.24}
\langle \widetilde{\mathcal E}({\bm k},s) \widetilde{\mathcal E}({\bm k}^{\prime},s^{\prime})\rangle_{H} &  \simeq&
\frac{4 \widetilde{V}^{2}}{N}  \zeta_{0}(\alpha) a_{33} (\alpha) \delta (s-s^{\prime}) \delta_{{\bm k},-{\bm k}^{\prime}} + \frac{ (d+2) \widetilde{V}^{2}}{N d^{2}}\, \delta_{{\bm k},-{\bm k}^{\prime}}  k^{2} \nonumber \\
&& \times \left\{ 1+ \frac{(d+8)}{2}\, a_{2}(\alpha) + \frac{2 d \zeta_{0}(\alpha) a_{33}(\alpha) \left[ 1+ 2 a_{2} (\alpha) \right]}{| \overline{\lambda}_{5}|-\zeta_{0}/2} \right\} e^{\overline{\lambda}_{5} |s-s^{\prime}|}, \nonumber \\
\end{eqnarray}
valid for $s,s^{\prime} \gg 1$. In this expression $a_{33}(\alpha)$ and $\overline{\lambda}_{5}$ are given  by Eqs.\ (\ref{apd.5}) and (\ref{apd.13}), respectively. The two main approximations made to derive the above expression are similar to those leading to Eq. (\ref{4.16}) and (\ref{4.18}). Firstly, the non-hydrodynamic component of particle velocity correlations is neglected once again. Secondly, it is written that
\begin{equation}
\label{5.25}
\Lambda^{+} ({\bm c}) {\bm \Sigma} ({\bm c}) \chi (c) \simeq \overline{\lambda}_{5} {\bm \Sigma} ({\bm c}) \chi (c),
\end{equation}
and $\overline{\lambda}_{5}$ is consistently obtained by means of Eq. (\ref{apd.12}).

When the expressions of thermal heat conductivity $\widetilde{\kappa}$ and the diffusive heat conductivity $\widetilde{\mu}$, Eqs.\ (\ref{5.16}) and (\ref{5.17}) respectively, are evaluated using the two above approximations, the results read
(see Appendix \ref{ApD})
\begin{equation}
\label{5.26}
\widetilde{\kappa} \simeq \frac {(d+2) \left[ 1+ 2 a_{2} (\alpha) \right]}{ 2 \left( 2 |\overline{\lambda}_{5}|-\zeta_{0} \right)},
\end{equation}
\begin{equation}
\label{5.27}
\widetilde{\mu} \simeq 2 \widetilde{\kappa} - \frac{ (d+2) \left[ 2+a_{2} (\alpha) \right] }{4 | \overline{\lambda}_{5}|}\, .
\end{equation}
The above expressions hold in the limit of large time $s$.  The above values for the thermal transport coefficients are indistinguishable of the results found in the first Sonine approximation \cite{BDKyS98,ByC01} for all values of $\alpha$. On the other hand, they are not equivalent to the expressions reported in ref. \cite{GSyM07}, although both are very close for $\alpha \gtrsim 0.65$.

Finally, to close the description provided by the fluctuating hydrodynamic equations derived here,  the correlation  between the fluctuating force appearing in the velocity equation and the fluctuating force in the energy equation is needed. It is verified in Appendix \ref{ApF} that
\begin{equation}
\label{5.28}
\langle \widetilde{\bm W} ({\bm k},s) \widetilde{\mathcal{E}}({\bm k}^{\prime},s^{\prime}) \rangle_{H}=0,
\end{equation}
as expected because of symmetry considerations.

\section{Final remarks}
\label{s6}
The objective here has been to derive fluctuating hydrodynamic equations for a dilute granular gas by extending methods which are familiar for normal gases. The main result obtained is the set of coupled  Langevin-like equations (\ref{3.4}), (\ref{4.14}), and (\ref{5.21}), describing the time evolution of the fluctuating hydrodynamic fields. These equations contain two kind of terms. There are terms which describe the hydrodynamic part of the
fluxes and of the cooling rate. The former have the same form as the macroscopic hydrodynamic fluxes, involving the Navier-Stokes transport coefficients. They have been made explicit in the equations. On the other hand, there appear the non-hydrodynamic components of the fluxes as well as some additional terms in the equation for the energy due to the dissipation in collisions. These latter terms have zero average and have been combined all together to define the noise terms in the Langevin-like equations. The auto-correlation functions of the noise terms in the velocity and energy equations are given in Eqs.\, (\ref{4.16}) and (\ref{5.24}), respectively, while the cross correlation function between both noise terms is indicated to vanish in Eq.\ (\ref{5.28}).

Generalizing fluctuating hydrodynamics to granular fluids entails several important differences from normal molecular fluids. Primary among them are the following.
\begin{enumerate}
\item The homogeneous reference state about which fluctuations are considered is not the Boltzmann equilibrium state, but the HCS. Its distribution function is not a simple function of the global invariants. Moreover, this reference state is time-dependent, although it is possible to consider a stationary representation of it by using an appropriate time scale (see Eq.\ (\ref{2.2})).
\item  Although the noise in the fluctuating inelastic Boltzmann equation is white, i.e. delta correlated in time, the noise terms in the equations for the velocity and energy fields have finite relaxation times. Moreover, the amplitude of their correlation functions is not determined by the Navier-Stokes transport coefficients, but involve new coefficients. In other words, the fluctuation-dissipation relations of the second kind \cite{KTyH85} are not verified.
\item The noise term in the equation for the energy has a zeroth order in the gradients contribution. This noise term is intrinsic to the inelasticity of collisions and has no analogue in molecular fluids.
 \end{enumerate}

The expressions for the two-time correlation functions of the noise terms have been computed using some approximations. It has been assumed that the velocity correlations between particles can be accurately approximated by their hydrodynamic part, identified as their projection onto the subspace of distributions generated by the hydrodynamic eigenfunctions of the linearized inelastic Boltzmann collision operator. The second approximation used consists in dealing with the hydrodynamic fluxes as if they were left eigenfunctions of the above mentioned linear operator. The two hypothesis can be justified on the basis of the following features: (1) they have been shown to lead to some predictions that are in very good agreement with molecular dynamics simulation results. This includes the fluctuations of the total energy of the system \cite{BGMyR04,VPBvWyT06} and also of the transversal component of the velocity field \cite{BMyG09}, (2) if the approximations are applied to the formal expressions of the Navier-Stokes transport coefficients, very accurate expressions are obtained as discussed in previous sections of this paper, and  (3) the second approximation mentioned above is an exact property in the case of the inelastic Maxwell model kinetic equation.

Some comments on the context and utility of the results in this work seem appropriate. The analysis has focussed on the fluctuations of the hydrodynamic fields in the HCS. In many experimental conditions of interest the system is far from a global homogeneous state. Nevertheless, the reference state studied here is relevant {\em locally} for more complex and realistic conditions, as it is the case of the equilibrium state in molecular fluids. For example, the transport coefficients such as the viscosity obtained here are the same functions of density and temperature as those in the associated nonlinear hydrodynamic equations applicable under more general conditions. Thus the context of relevance of the equations derived in the present work are expected to transcend the limitations associated with the state considered and extend to states for which the nonlinear Navier-Stokes are required to characterize the macroscopic hydrodynamic fields.

\section{Acknowledgements}

This research was supported by the Ministerio de Educaci\'{o}n y
Ciencia (Spain) through Grant No. FIS2008-01339 (partially financed
by FEDER funds).

\appendix
\section{Dimensionless linear Boltzmann collision operator}
\label{ApA}
The dimensionless binary collision operator $\overline{T}_{0}({\bm c}_{1},{\bm c}_{2})$ for inelastic hard spheres or disks
is defined by
\begin{equation}
\label{apa.1}
\overline{T}_{0}({\bm c}_{1},{\bm c}_{2})=  \int d
\widehat{\bm \sigma}\, \Theta ({\bm c}_{12} \cdot
\widehat{\bm \sigma}) {\bm c}_{12} \cdot \widehat{\bm
\sigma}   \left[ \alpha^{-2} b_{\bm \sigma}^{-1}({\bm c}_{1},{\bm
c}_{2}) -1 \right],
\end{equation}
where ${\bm c}_{12} \equiv {\bm c}_{1}- {\bm c}_{2}$, $d
\widehat{\bm \sigma}$ is the solid angle element for the unit vector $\widehat{\bm \sigma}$,
$\Theta$ is the Heaviside step function, and $  b_{\bm \sigma}^{-1}({\bm c}_{1},{\bm
c}_{2})$ is an operator changing all the functions of ${\bm c}_{1}$ and ${\bm c}_{2}$ to its right by the
same functions of the precollisonal velocities ${\bm c}^{*}_{1}$ and ${\bm c}^{*}_{2}$, given by
\begin{eqnarray}
\label{apa.2} {\bm c}^{*}_{1} \equiv b_{\bm \sigma}^{-1} {\bm c}_{1}=
{\bm c}_{1}-\frac{1+\alpha}{2 \alpha} ( \widehat{\bm \sigma} \cdot
{\bm c}_{12} ) \widehat{\bm \sigma},
\nonumber \\
{\bm c}^{*}_{2} \equiv b_{\bm \sigma}^{-1} {\bm c}_{2}= {\bm
c}_{2}+\frac{1+\alpha}{2 \alpha} ( \widehat{\bm \sigma} \cdot {\bm
c}_{12} ) \widehat{\bm \sigma}.
\end{eqnarray}
The expression of the linearized Boltzmann collision operator  $\Lambda ({\bm c})$ is \cite{BDyR03}
\begin{equation}
\label{apa.3}
\Lambda({\bm c}_{1}) \equiv \int d {\bm c}_{2}\,
\overline{T}_{0}({\bm c}_{1},{\bm c}_{2}) (1+P_{12}) \chi
({c}_{2})-\frac{\zeta_{0}}{2} \frac{\partial}{\partial {\bm c}_{1}}
\cdot {\bm c}_{1}.
\end{equation}
The operator $P_{12}$ interchanges the labels of particles $1$ and
$2$ of the quantities to its right, $\chi (c)$ is the scaled velocity distribution
of the HCS defined in Eq.\ (\ref{2.12}), and $\zeta_{0}$ is the dimensionless cooling rate for the
decay of the temperature of the HCS in the time scale $s$,
\begin{equation}
\label{apa.4}
\frac{dT_{H}(s)}{ds} = - \zeta_{0} T_{H}(s),
\end{equation}
\begin{equation}
\label{apa.5}
 \zeta_{0}=\frac{(1-\alpha^{2})\pi^{\frac{d-1}{2}}}{2\,
\Gamma \left( \frac{d+3}{2} \right)d} \int d{\bm c}_{1} \int d{\bm
c}_{2}\, c_{12}^{3}\chi({c}_{1}) \chi({c}_{2}).
\end{equation}
Approximated expressions for the distribution function of the HCS and for the cooling rate
have been obtained by expanding the functions in Sonine polynomials and keeping only the
lowest orders \cite{GyS95,vNyE98}. The distribution function has the form
\begin{equation}
\label{apa.6}
\chi(c)= \frac{e^{-c^{2}}}{\pi^{d/2}}\, \left[ 1
+a_{2}(\alpha) S^{(2)} (c^{2}) \right],
\end{equation}
where
\begin{equation}
\label{apa.7}
S^{(2)}(c^{2})= \frac{c^{4}}{2}-\frac{d+2}{2}\, c^{2} +\frac{d(d+2)}{8}
\end{equation}
and
\begin{equation}
\label{apa.8}
a_{2}(\alpha)= \frac{16(1-\alpha)(1-2 \alpha^{2})}{9+24d+(8d-41)\alpha+30 \alpha^{2}
-30 \alpha^{3}}\, .
\end{equation}
Equation (\ref{apa.6}) is used all along this paper to carry out explicit calculations. The approximate expression for the cooling rate is:
\begin{equation}
\label{apa.9}
\zeta_{0}= \frac{ \sqrt{2} \pi^{(d-1)/2} (1-\alpha^{2})}{\Gamma \left(d/2 \right) d} \left[ 1+ \frac{3 a_{2}(\alpha)}{16} \right].
\end{equation}

\section{Derivation of Eq.\ (\protect{\ref{4.10}})}
\label{ApB}
From Eq. (\ref{4.4}) and using Eq.\ (\ref{2.16}), it follows that for small wavevector ${\bm k}$ it is
\begin{eqnarray}
\label{apb.1}
\langle {\sf R}({\bm k},s) {\sf R}({\bm k}^{\prime},s^{\prime})\rangle_{H} &=& \frac{\widetilde{V}^{2}}{N}\, \delta_{{\bm k},-{\bm k}^{\prime}} \int d{\bm c}_{1} \int d{\bm c}_{2}\, {\sf \Delta} ({\bm c}_{1}) {\sf \Delta} ({\bm c}_{2}) \nonumber \\ && \times \int_{0}^{s} ds_{1}  \int_{0}^{s^{\prime}} ds_{2}\, e^{s_{1} \Lambda ({\bm c}_{1}) + s_{2} \Lambda ({\bm c}_{2})}
\delta (s-s^{\prime}-s_{1}+s_{2}) \mathcal{P}_{\perp}^{(1)} \mathcal{P}_{\perp}^{(2)} \widetilde{\Gamma} ({\bm c}_{1}, {\bm c}_{2} ), \nonumber \\
\end{eqnarray}
where it has been used that to lowest order in $k$ the inhomogeneous linear Boltzmann operator $\Lambda({\bm k},{\bm c})$ can be replaced by the homogeneous one, $\Lambda ({\bm c})$. The projection operators $\mathcal{P}_{\perp}^{(1)}$ and $\mathcal{P}_{\perp}^{(2)}$ act on functions of ${\bm c}_{1}$ and ${\bm c}_{2}$, respectively. Suppose that $s^{\prime}  >s$. The integration over $s_{2}$ can be easily carried out to get
\begin{eqnarray}
\label{apb.2}
\langle {\sf R}({\bm k},s) {\sf R}({\bm k}^{\prime},s^{\prime})\rangle_{H} &=& \frac{\widetilde{V}^{2}}{N}\, \delta_{{\bm k},-{\bm k}^{\prime}} \int d{\bm c}_{1} \int d{\bm c}_{2}\, {\sf \Delta} ({\bm c}_{1}) {\sf \Delta} ({\bm c}_{2}) \nonumber \\ && \times e^{(s^{\prime}-s) \Lambda ({\bm c}_{2})} \int_{0}^{s} ds_{1}\, e^{s_{1} \left[ \Lambda ({\bm c}_{1})+ \Lambda ({\bm c}_{2})\right]} \mathcal{P}_{\perp}^{(1)} \mathcal{P}_{\perp}^{(2)} \widetilde{\Gamma} ({\bm c}_{1}, {\bm c}_{2} ).
\end{eqnarray}
For $s^{\prime} >s\gg 1$ and assuming that all the non-hydrodynamic components of $\widetilde{\Gamma} ({\bm c}_{1}, {\bm c}_{2})$ correspond to negative eigenvalues of $\Lambda ({\bm c}_{1})$ and $\Lambda ({\bm c}_{2})$, the above relation can be simplified to
\begin{equation}
\label{apb.3}
 \langle {\sf R}({\bm k},s) {\sf R}({\bm k}^{\prime},s^{\prime})\rangle_{H} = \frac{\widetilde{V}^{2}}{N}\, \delta_{{\bm k},-{\bm k}^{\prime}} \int d{\bm c}_{1} \int d{\bm c}_{2}\, {\sf \Delta} ({\bm c}_{1}) {\sf \Delta} ({\bm c}_{2})  e^{(s^{\prime}-s) \Lambda ({\bm c}_{2})} \widetilde{\phi}_{H}({\bm c}_{1},{\bm c}_{2}),
\end{equation}
where $\widetilde{\phi}_{H}({\bm c}_{1},{\bm c}_{2})$ obeys Eq.\ (\ref{4.12}). The symmetry of the tensor ${\sf \Delta}({\bm c})$, the isotropy of the operator $\Lambda ({\bm c})$ and the invariance of $\widetilde{\Gamma} ({\bm c}_{1}, {\bm c}_{2})$ under rotations of ${\bm c}_{1}$ and ${\bm c}_{2}$ imply that the right hand side of Eq. (\ref{apb.3}) must have the form given in Eq. (\ref{4.10}). The case $s>s^{\prime} \gg 1$ follows trivially.

\section{Approximated evaluation of the function $G(s)$ defined in Eq. (\protect{\ref{4.11}})}
\label{ApC}
Using the approximation given in Eq.\ (\ref{4.13}) into Eq. (\ref{4.11}) gives
\begin{equation}
\label{apc.1}
G(s) \simeq \frac{1}{d^{2}+d-2} \sum_{i,j}^{d} \int d{\bm c}\,  \Delta_{ij} ({\bm c})
 e^{s \Lambda ({\bm c})} \Delta_{ij} ({\bm c}) \chi (c),
\end{equation}
and taking into account the symmetry of the operator $\Lambda ({\bm c})$ and of the function $\chi (c)$, this expression is seen to be equivalent to
\begin{eqnarray}
\label{apc.2}
G(s)& = & \int d{\bm c}\, \Delta_{xy} ({\bm c})e^{s \Lambda ({\bm c})} \Delta_{ij} ({\bm c}) \chi (c) \nonumber \\
&=& \langle \Delta_{xy} \chi | e^{s \Lambda } \Delta_{xy} \chi \rangle \nonumber \\
& = & \langle e^{s \Lambda^{+}} \Delta_{xy} \chi | \Delta_{xy} \chi  \rangle,
\end{eqnarray}
where $\Lambda^{+}({\bm c})$ is the adjoint of $\Lambda({\bm c})$ defined by
\begin{equation}
\label{apc.3}
 \langle g | \Lambda h \rangle  = \langle \Lambda^{+} g | h \rangle^{*},
\end{equation}
for arbitrary functions $g({\bm c}) $ and $h({\bm c})$ of the Hilbert space. Now the approximation is made that $\Delta_{xy} ({\bm c}) \chi(c)$ is an eigenfunction of $\Lambda^{+}$, being $\overline{\lambda}_{4}$ the eigenvalue, as expressed by Eq.\ (\ref{4.17}). This approximation is prompted by the fact that it is exact for the inelastic Maxwell model of granular gases \cite{BGyM10}. Use of Eq.\ (\ref{4.17}) into Eq.\ (\ref{apc.2}) yields
\begin{equation}
\label{apc.4}
G(s) = e^{s \overline{\lambda}_{4}} \int d{\bm c}\, c_{x}^{2} c_{y}^{2} \chi (c) = \frac{1+ a_{2}(\alpha)}{4} \, e^{ s \overline{\lambda}_{4}},
\end{equation}
where the expression of $\chi (c)$ in the first Sonine approximation, Eq.\ (\ref{apa.6}) has been employed to evaluate the velocity integral.

To determine $\overline{\lambda}_{4}$, Eq.\ (\ref{4.17}) is multiplied by $\Delta_{xy}({\bm c})$ and integrated over ${\bm c}$ to get
\begin{equation}
\label{apc.5}
\overline{\lambda}_{4} = \frac{4}{1+ a_{2} (\alpha)}\, \int d{\bm c}\,  c_{x} c_{y} \Lambda^{+}  ({\bm c}) \Delta_{xy}({\bm c}) \chi (c).
\end{equation}
The evaluation of the velocity integral on the right hand side of the above expression using once again Eq.\ (\ref{apa.6}) is a lengthly but quite standard calculation. The result is given in Eq.\ (\ref{4.19}), where
\begin{equation}
\label{apc.6}
I(\alpha)= - \frac{(2d+3-3 \alpha)(1+\alpha) \pi^{\frac{d-1}{2}}}{2 \sqrt{2} d (d+2) \Gamma (d/2)}\, \left[ 1+ \frac{23 a_{2}(\alpha)}{16} \right].
\end{equation}
In the same approximation as introduced above, the expression for the shear viscosity, $\widetilde{\eta} (s)$, given by Eq.\ (\ref{4.7}) in the limit of large $s$ becomes
\begin{eqnarray}
\label{apc.7}
\widetilde{\eta} & \simeq & \int_{0}^{\infty} ds^{\prime} \int  d{\bm c}\, \Delta_{xy} ({\bm c}) e^{s^{\prime} \left[ \Lambda ({\bm c}) - \zeta_{0}/2 \right]}  \xi_{2,x} ({\bm c}) c_{y} \nonumber \\
& = & \int_{0}^{\infty} d s^{\prime}\ e^{-s^{\prime} \zeta_{0}/2} \langle e^{s^{\prime} \Lambda^{+}} \left ( \chi \Delta_{xy} \right) | \xi_{2,x}  c_{y} \rangle \nonumber \\
& \simeq &   \langle  \chi \Delta_{xy}  | \xi_{2,x} ({\bm c}) c_{y} \rangle \int_{0}^{\infty} ds^{\prime}\, e^{-s^{\prime} \left( \zeta_{0}/2 - \overline{\lambda}_{4} \right)} \nonumber \\
&=& -  \langle \chi \Delta_{xy} | \xi_{2,x}  c_{y} \rangle
\left( \overline{\lambda}_{4} - \frac{\zeta_{0}}{2} \right)^{-1}.
\end{eqnarray}
Taking into account that
\begin{equation}
\label{apc.8}
 \langle \chi \Delta_{xy} | \xi_{2,x}  c_{y} \rangle = \frac{1}{2},
\end{equation}
Eq. (\ref{4.20}) is obtained.

\section{Correlation of the noise term in the equation for the energy field}
\label{ApD}
In the following calculations, it will be assumed without loss of generality that $s> s^{\prime}$. From Eq.
(\ref{5.22}) and keeping only contributions up to order $k^{2}$ it is obtained that
\begin{eqnarray}
\label{apd.1}
\langle  \widetilde{\mathcal E}({\bm k},s) \widetilde{\mathcal E}({\bm k}^{\prime},s^{\prime})\rangle_{H}
& \simeq & \langle  \widetilde{S}_{\epsilon} ({\bm k},s)\widetilde{S}_{\epsilon} ({\bm k}^{\prime},s^{\prime})\rangle_{H}
- \frac{4}{d^{2}} {\bm k} \cdot \langle   {\bm Z} ({\bm k},s) {\bm Z} ({\bm k}^{\prime},s^{\prime})\rangle_{H} \cdot {\bm k}^{\prime} \nonumber \\
&& - \frac{2i}{d} {\bm k} \cdot \langle  \delta_{1}^{(I)} {\bm \phi} ({\bm k},s )\widetilde{S}_{\epsilon} ({\bm k}^{\prime},s^{\prime})\rangle_{H} - \frac{2i}{d} {\bm k} \cdot \langle  {\bm Z} ({\bm k},s) \widetilde{S}_{\epsilon} ({\bm k}^{\prime},s^{\prime})\rangle_{H}. \nonumber \\
\end{eqnarray}
Upon writing the above equation, it has been taken into account that $\delta_{1}^{(I)} {\bm \phi} ({\bm k},s )$ is at least of first order in $k$, as it is directly realized from its expression in Eq.\ (\ref{5.20}). Moreover, it has been used that
\begin{equation}
\label{apd.2}
\langle  \widetilde{S}_{\epsilon} ({\bm k},s) \delta {\bm \phi} ({\bm k}^{\prime},s^{\prime})\rangle_{H}=0,
\end{equation}
as a consequence of Eq.\ (\ref{3.15}). Consider first the self-correlation of the intrinsic noise $\widetilde{S}_{\epsilon}$, given by Eq.\ (\ref{3.14}) or, equivalently, by
\begin{equation}
\label{apd.3}
\langle  \widetilde{S}_{\epsilon}({\bm k},s)
\widetilde{S}_{\epsilon}({\bm
k}^{\prime},s^{\prime})\rangle_{\text{H}} = \frac{4 \widetilde{V}^{2}}{ N} \delta_{{\bm k},-{\bm k}^{\prime}}
\delta( s- s^{\prime}) \int d{\bm c} \int d {\bm c}{^\prime}\, \overline{\xi}_{3}({\bm c}) \overline{\xi}_{3}({\bm c}^{\prime}) \widetilde{\Gamma} ({\bm c}, {\bm c}^{\prime}).
\end{equation}
The velocity integral appearing on the right hand side of this equation can be evaluated exactly using the expression for $\chi (c)$  in the first Sonine approximation. Nevertheless, for the sake of consistency, here the non-hydrodynamic components of the particle velocity correlations in the HCS will be neglected, as it was done when solving Eq.\, (\ref{4.12}). With this approximation, it was shown in ref. \cite{BGMyR04} that
\begin{equation}
\label{apd.4}
\langle  \widetilde{S}_{\epsilon}({\bm k},s)
\widetilde{S}_{\epsilon}({\bm
k}^{\prime},s^{\prime})\rangle_{\text{H}} \simeq  \frac{4 \widetilde{V}^{2}}{ N} \zeta_{0} a_{33}(\alpha) \delta_{{\bm k},-{\bm k}^{\prime}} \delta( s- s^{\prime}),
\end{equation}
with
\begin{equation}
\label{apd.5}
a_{33}(\alpha)=\frac{d+1}{2d}+\frac{d+2}{4d}\,
a_{2}(\alpha)+ b(\alpha),
\end{equation}
\begin{equation}
\label{apd.6}
b(\alpha)= \frac{2+d-6d^{2}-(10-15d +2d^{2})
\alpha-2(2+7d)\alpha^{2}+2(10-d)\alpha^{3}}{6d(2d+1)-2d(11-2d)\alpha+12
d \alpha^{2}-12 d \alpha^{3}} \, .
\end{equation}

Consider next the self-correlation of  ${\bm Z} ({\bm k},s)$. To the lowest order in $k$, it is
\begin{eqnarray}
\label{apd.7}
{\bm k} \cdot \langle   {\bm Z} ({\bm k},s) {\bm Z} ({\bm k}^{\prime},s^{\prime})\rangle_{H} \cdot {\bm k}^{\prime} & \simeq & \int d{\bm c} \int d{\bm c}{^\prime}\, {\bm k} \cdot {\bm \Sigma} ({\bm c}) {\bm k}^{\prime} \cdot {\bm \Sigma} ({\bm c}^{\prime}) \int_{0}^{s} ds_{1} \int _{0}^{s^{\prime}} ds_{2}\, \nonumber \\
&& \times e^{s_{1} \Lambda ({\bm c}) + s_{2} \Lambda ({\bm c}^{\prime})} \langle  \widetilde{S} ({\bm k},{\bm c},s-s_{1}) \widetilde{S} ({\bm k},{\bm c}^{\prime},s^{\prime}-s_{2})\rangle_{H} \nonumber \\
&=& - \frac{\widetilde{V}^{2}}{Nd} \delta_{{\bm k},-{\bm k}^{\prime}}  k^{2} \int d{\bm c} \int d{\bm c}^{\prime}\,
{\bm \Sigma}({\bm c}) \cdot {\bm \Sigma}({\bm c}^{\prime}) \nonumber \\
& & \times \int_{0}^{s^{\prime}} ds_{2}\, e^{s_{2} \Lambda ({\bm c}^{\prime})} e^{(s+s_{2}-s^{\prime}) \Lambda ({\bm c})}
\widetilde{\Gamma} ({\bm c},{\bm c}^{\prime}).
\end{eqnarray}
For $s>s^{\prime} \gg 1$, and assuming again that the HCS is stable with respect to homogeneous perturbations of the velocity, so that all the non-hydrodynamic eigenvalues  of $\Lambda ({\bm c})$ must be negative, carrying out the integration over $s_{2}$ yields
\begin{eqnarray}
\label{apd.8}
{\bm k} \cdot \langle   {\bm Z} ({\bm k},s) {\bm Z} ({\bm k}^{\prime},s^{\prime})\rangle_{H} \cdot {\bm k}^{\prime} &=& - \frac{\widetilde{V}^{2}}{Nd} \delta_{{\bm k},-{\bm k}^{\prime}}  k^{2} \int d{\bm c} \int d{\bm c}^{\prime}\,
{\bm \Sigma}({\bm c}) \cdot {\bm \Sigma}({\bm c}^{\prime}) \nonumber \\
&& \times e^{(s-s^{\prime}) \Lambda ({\bm c})} \widetilde \phi_{H} ({\bm c},{\bm c}^{\prime}),
\end{eqnarray}
where $\widetilde{\phi}_{H}$ is defined in Eq.\ (\ref{4.12}). The right hand side of the above relation can be evaluated by means of an approximation scheme similar to that used in Sec. \ref{s4} to compute $G(s)$. Since the justification of the approximations to be made is the same as discussed in Sec.\ \ref{s4}, it will not be repeated here. To begin with, Eq.\ (\ref{4.13}) is used to write
\begin{equation}
\label{apd.9}
{\bm k} \cdot \langle   {\bm Z} ({\bm k},s) {\bm Z} ({\bm k}^{\prime},s^{\prime})\rangle_{H} \cdot {\bm k}^{\prime}  \simeq
- \frac{\widetilde{V}^{2}}{Nd} \delta_{{\bm k},-{\bm k}^{\prime}}  k^{2} \int d{\bm c}\, {\bm \Sigma} ({\bm c}) \cdot
e^{(s-s^{\prime}) \Lambda ({\bm c})} {\bm \Sigma}({\bm c}) \chi (c).
\end{equation}
Now it is assumed that (compare with Eq.\ (\ref{4.17}))
\begin{equation}
\label{apd.10}
\Lambda^{+} ({\bm c}) {\bm \Sigma} ({\bm c}) \chi (c) \simeq \overline{\lambda}_{5} {\bm \Sigma} ({\bm c}) \chi (c),
\end{equation}
so that Eq. (\ref{apd.9}) is approximated by
\begin{eqnarray}
\label{apd.11}
{\bm k} \cdot \langle  {\bm Z} ({\bm k},s) {\bm Z} ({\bm k}^{\prime},s^{\prime})\rangle_{H} \cdot {\bm k}^{\prime} & \simeq &
- \frac{\widetilde{V}^{2}}{Nd} \delta_{{\bm k},-{\bm k}^{\prime}}  k^{2} e^{(s-s^{\prime}) \overline{\lambda}_{5}}
\int d{\bm c}\, {\bm \Sigma}^{2}({\bm c}) \chi ( c) \nonumber \\
& \simeq &  - \frac{\widetilde{V}^{2}}{8 N} \delta_{{\bm k},-{\bm k}^{\prime}}  k^{2} (d+2) \left[ 2 + (d+8) a_{2} (\alpha) \right] e^{(s-s^{\prime}) \overline{\lambda}_{5}}.
\end{eqnarray}
To determine the value of $\overline{\lambda}_{5}$, Eq.\ (\ref{apd.10}) is multiplied by $c_{x}$ and afterwards integrated over the velocity ${\bm c}$, yielding
\begin{equation}
\label{apd.12}
\overline{\lambda}_{5} = \frac{1}{(d+2) a_{2}(\alpha)}\, \int d{\bm c}\, c_{x} \Lambda^{+}({\bm c}) \Sigma_{x} ({\bm c}) \chi (c).
\end{equation}
From here it is obtained
\begin{equation}
\label{apd.13}
\overline{\lambda}_{5} = \frac{ 4 J(\alpha)}{(d+2) a_{2} (\alpha)} + \frac{\zeta_{0}(\alpha)}{a_{2}(\alpha)} + \frac{3 \zeta_{0} (\alpha)}{2}\,  ,
\end{equation}
\begin{eqnarray}
\label{apd.14}
J(\alpha) & = & - \frac{\pi^{(d-1)/2}(1+\alpha)}{32 \sqrt{2} d \Gamma \left(d/2 \right)} \nonumber \\
 && \times \left\{16(2+d) (1-\alpha)+a_{2}(\alpha) \left[ 70+47d -3(34+5d) \alpha \right] \right\}.
\end{eqnarray}

The third term on the right hand side of Eq.\ (\ref{apd.1}) is evaluated as follows. It is
\begin{eqnarray}
\label{apd.15}
{\bm k} \cdot \langle  \delta_{1}^{(I)} {\bm \phi} ({\bm k},s )\widetilde{S}_{\epsilon} ({\bm k}^{\prime},s^{\prime})\rangle_{H}
 & = & \frac{1}{2} \int_{0}^{s} ds_{1} \int d{\bm c}\, {\bm k} \cdot {\bm \Sigma} ({\bm c}) e^{(s-s_{1}) \Lambda ({\bm c})}
i {\bm k} \cdot {\bm c} e^{-\zeta_{0} s_{1}/2} \nonumber \\
&& \times \int_{s_{1}}^{s} ds_{2}\, e^{\zeta_{0} s_{2}/2}  \langle \widetilde{S}_{\epsilon} ({\bm k},s_{2})\widetilde{S}_{\epsilon} ({\bm k}^{\prime},s^{\prime})\rangle_{H} \xi_{3}({\bm c}).
\end{eqnarray}
Using the approximation defined in Eq.\ (\ref{apd.10}), it is
\begin{eqnarray}
\label{apd.16}
\int d{\bm c}\, {\bm \Sigma} ({\bm c})  e^{(s-s_{1}) \Lambda ({\bm c})} {\bm c} \xi_{3}({\bm c})
& \simeq & e^{(s-s_{1}) \overline{\lambda}_{5}} \int d{\bm c}\, {\bm \Sigma}({\bm c}) {\bm c} \xi_{3}({\bm c})
\nonumber \\
&=& e^{(s-s_{1}) \overline{\lambda}_{5}} \frac{d+2}{2} \left[ 1 + 2 a_{2} (\alpha) \right] {\sf I}.
\end{eqnarray}
Then, use of this result and Eq. (\ref{apd.4}) into Eq.\ (\ref{apd.15}) yields
\begin{eqnarray}
\label{apd.17}
{\bm k} \cdot \langle  \delta_{1}^{(I)} {\bm \phi} ({\bm k},s )\widetilde{S}_{\epsilon} ({\bm k}^{\prime},s^{\prime})\rangle_{H}
   & \simeq &   \frac{i (d+2) \left[1+2 a_{2} (\alpha) \right] \zeta_{0}(\alpha) a_{33} (\alpha) \widetilde{V}^{2} k^{2}}{N}\, \nonumber \\
 && \times  \delta_{{\bm k}, -{\bm k}^{\prime}} e^{s^{\prime}\,  \zeta_{0}/2} e^{s \overline{\lambda}_{5}} \int_{0}^{s^{\prime}} ds_{1}\, e^{-s_{1} \left( \overline{\lambda}_{5} + \zeta_{0}/2 \right)}.
\end{eqnarray}
In the limit $s \geq s^{\prime} \gg 1$, the above result reduces to
\begin{equation}
\label{apd.18}
{\bm k} \cdot \langle  \delta_{1}^{(I)} {\bm \phi} ({\bm k},s )\widetilde{S}_{\epsilon} ({\bm k}^{\prime},s^{\prime})\rangle_{H}
\simeq    \frac{i (d+2) \left[1+2 a_{2} (\alpha) \right] \zeta_{0}(\alpha) a_{33} (\alpha) \widetilde{V}^{2} k^{2}\delta_{{\bm k}, -{\bm k}^{\prime}} e^{\overline{\lambda}_{5} (s-s^{\prime})}}{N \left(|\overline{\lambda}_{5}|
- \zeta_{0}/2 \right)}.
\end{equation}
The calculation of the last term on the right hand side of Eq.\ (\ref{apd.1}) is much more involved. Nevertheless, in the next appendix some evidence is presented yielding the conclusion that it can be safely neglected since it leads to a much smaller contribution than the other terms, at least for not too strong inelasticity. Then, putting  all the results obtained in this appendix together, Eq. (\ref{5.24}) follows.

The long time limit of the expression for the thermal heat conductivity $\widetilde{\kappa}$, Eq. (\ref{5.16}), is easily evaluated within the approximation scheme developed above,
\begin{eqnarray}
\label{apd.19}
\widetilde{\kappa} &\simeq& \frac{1}{2d} \int_{0}^{\infty} ds^{\prime} \int d{\bm c}\, e^{s^{\prime} \left( \overline{\lambda}_{5} + \zeta_{0}/2 \right)} \xi_{3}({\bm c}) {\bm \Sigma} ({\bm c}) \cdot {\bm c}
\nonumber \\
&=& - \frac{1}{2d} \left( \overline{\lambda}_{5} + \frac{\zeta_{0}}{2} \right)^{-1}
\int d{\bm c}\, \xi_{3}({\bm c}) {\bm \Sigma} ({\bm c}) \cdot {\bm c}.
\end{eqnarray}
By using now that
\begin{equation}
\label{apd.20}
\int d{\bm c}\, \xi_{3}({\bm c}) {\bm \Sigma} ({\bm c}) \cdot {\bm c} = \frac{d(d+2)}{2} \left[ 1 + 2 a_{2} (\alpha)
\right] ,
\end{equation}
Eq. (\ref{apd.19}) yields Eq. (\ref{5.26}). Equation (\ref{5.27}) is derived in a similar way.

\section{Analysis of the last term on the right hand side of Eq.\ (\ref{apd.1})}
\label{ApE}
\begin{figure}
\includegraphics[scale=0.7,angle=0]{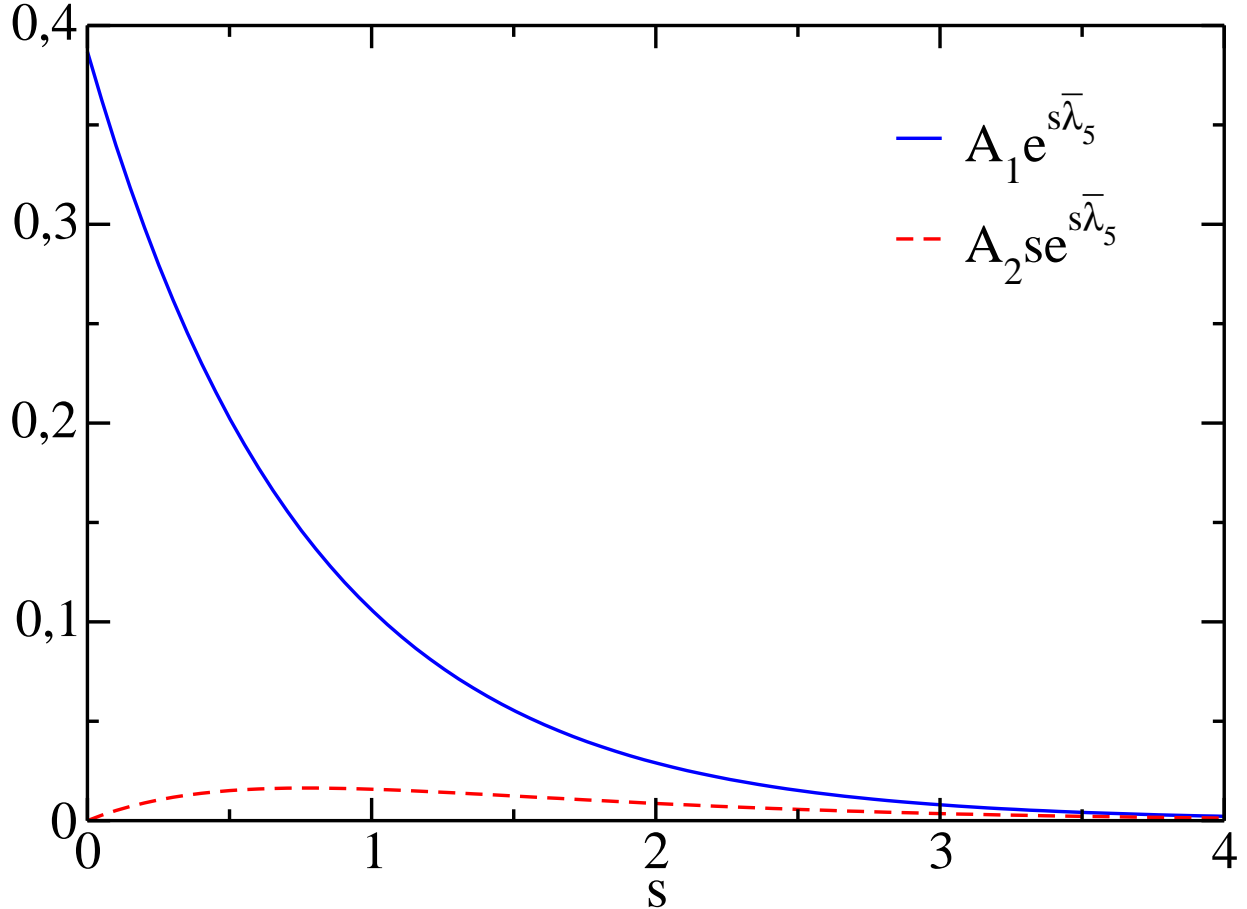}
\caption{(Color online) Comparison of the dimensionless quantities $A_{1}e^{s \overline{\lambda}_{5}}$ (solid blue line) and $A_{2}s e^{s \overline{\lambda}_{5}}$ (dashed red line) as a function of the dimensionless time $s$ for $\alpha=0.6$. The definitions of $A_{1}$ and $A_{2}$ are given in Eqs.\ (\protect{\ref{ape.8}}) and (\protect{\ref{ape.9}}) and the case $d=2$ has been considered. It is seen that the contribution $K_{1}(s)$ associated to the amplitud $A_{2}$, given in Eq.\ (\protect{\ref{ape.3}}) can be accurately neglected.
\label{fig1}}
\end{figure}

\begin{figure}
\includegraphics[scale=0.7,angle=0]{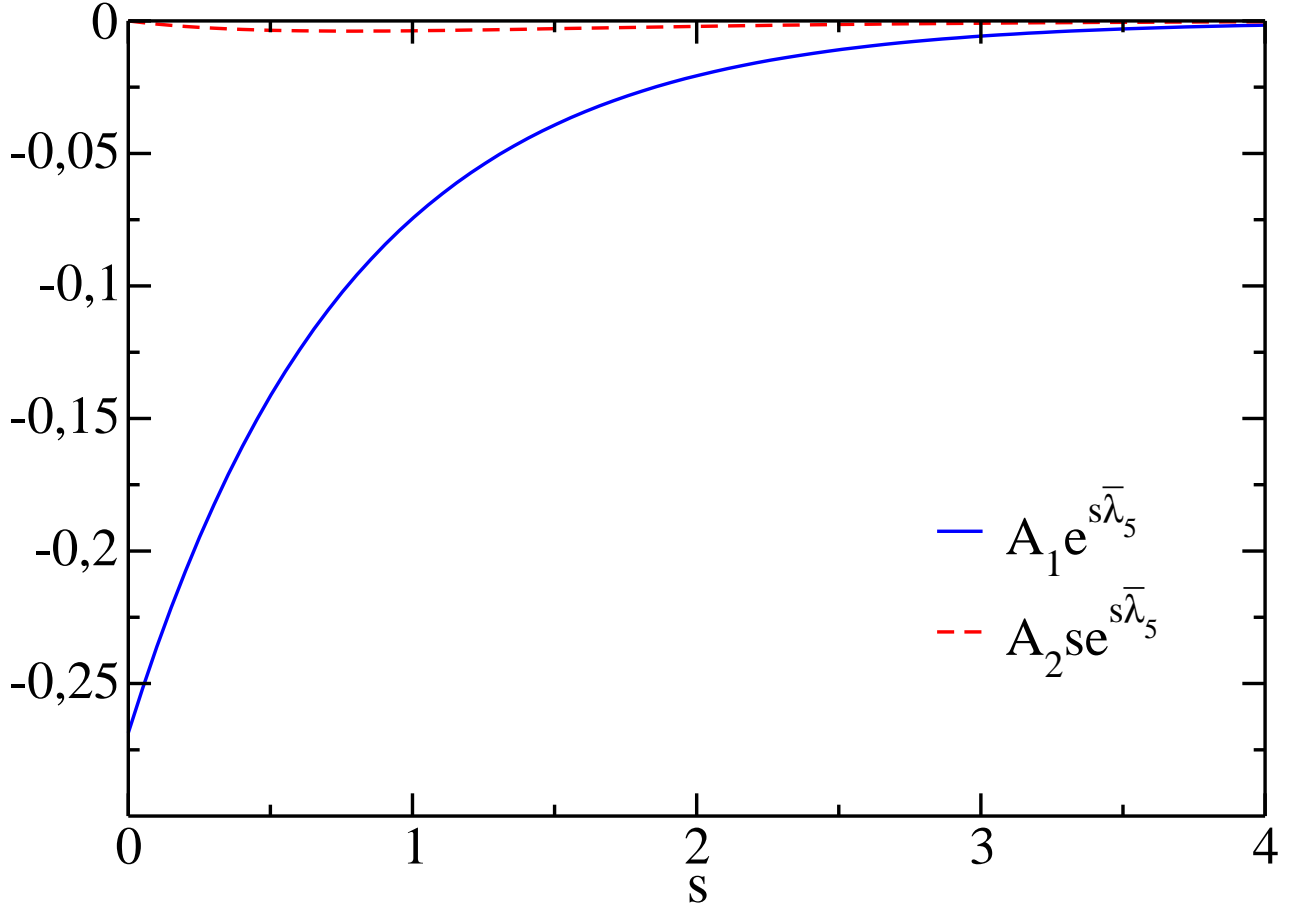}
\caption{(Color online) The same as in Fig.\ \protect{\ref{fig1}}, but for $\alpha=0.9$.
\label{fig2}}
\end{figure}

It is
\begin{eqnarray}
\label{ape.1}
-\frac{2i}{d}{\bm k} \cdot \langle  {\bm Z} ({\bm k},s) \widetilde{S}_{\epsilon} ({\bm k}^{\prime},s^{\prime})\rangle_{H} &=& - \frac{4i}{d^{2}}\, {\bm k} \cdot \int_{0}^{s} d s_{1} \int d{\bm c} \int d{\bm c}^{\prime}\, c^{\prime 2} {\bm \Sigma} ({\bm c})
\mathcal{U} ({\bm k}, {\bm c},s_{1}) \nonumber \\
&& \times \mathcal{P}_{\perp} \langle  \widetilde{S}({\bm k},{\bm c},s-s_{1})
\widetilde{S} ({\bm k}^{\prime}, {\bm c}^{\prime}, s^{\prime})\rangle_{H} \nonumber \\
&=& -\frac{4 i \widetilde{V}^{2}}{Nd^{2}} \delta_{{\bm k},-{\bm k}^{\prime}} {\bm k} \cdot \int d{\bm c} \int d{\bm c}^{\prime}\, c^{\prime 2} {\bm \Sigma} ({\bm c}) \mathcal{U} ({\bm k}, {\bm c},s-s^{\prime}) \mathcal{P}_{\perp} \widetilde{\Gamma} ({\bm c},{\bm c}^{\prime}) \nonumber \\
& \simeq &  \frac{4i \widetilde{V}^{2}}{Nd^{2}} \delta_{{\bm k},-{\bm k}^{\prime}} {\bm k} \cdot \int d{\bm c} \int d{\bm c}^{\prime}\, c^{\prime 2} {\bm \Sigma} ({\bm c}) \mathcal{U} ({\bm k}, {\bm c},s-s^{\prime}) \nonumber \\
&& \times \mathcal{P}_{\perp} \left[ \Lambda ({\bm c}^{\prime})+ \Lambda ({\bm c}) \right] \chi(c) \delta \left( {\bm c}-{\bm c}^{\prime} \right).
\end{eqnarray}
In the last transformation, the contribution to $\mathcal{P}_{\perp} \widetilde{\Gamma} ({\bm c},{\bm c}^{\prime})$ due to the velocity correlations in the HCS has been neglected, as done everywhere along this paper. Therefore,
\begin{equation}
\label{ape.2}
- \frac{2i}{d}{\bm k} \cdot \langle  {\bm Z} ({\bm k},s) \widetilde{S}_{\epsilon} ({\bm k}^{\prime},s^{\prime})\rangle_{H} = K_{1}(s-s^{\prime}) + K_{2}(s-s^{\prime}),
\end{equation}
with
\begin{eqnarray}
\label{ape.3}
 K_{1}(s) & =  &\frac{4 i \widetilde{V}^{2}}{Nd^{2}} \delta_{{\bm k},-{\bm k}^{\prime}} {\bm k} \cdot \int d{\bm c}
 \int d{\bm c}^{\prime}\, c^{\prime 2}{\bm \Sigma} ({\bm c}) \mathcal{U} ({\bm k}, {\bm c},s) \nonumber \\
 && \times \mathcal{P}_{\perp} \Lambda ({\bm c}^{\prime}) \chi (c) \delta ({\bm c}-{\bm c}^{\prime}),
\end{eqnarray}
\begin{eqnarray}
\label{ape.4}
K_{2}(s) & = &\frac{4i \widetilde{V}^{2}}{Nd^{2}} \delta_{{\bm k},-{\bm k}^{\prime}} {\bm k} \cdot \int d{\bm c}\, {\bm \Sigma} ({\bm c})\, \mathcal{U} ({\bm k}, {\bm c},s) \nonumber \\
 && \times \mathcal{P}_{\perp} \Lambda ({\bm c}) \chi (c) c^{2}.
\end{eqnarray}
Using again the approximation in Eq.\ (\ref{5.1}), the definition of $K_{1}(s)$ in Eq.\ (\ref{ape.3}) can be easily rewritten as
\begin{equation}
\label{ape.5}
K_{1}(s) \simeq -\frac{2i\widetilde{V}^{2} \zeta_{0}}{Nd}\, \delta_{{\bm k},-{\bm k}^{\prime}} {\bm k} \cdot \int d{\bm c}\, {\bm \Sigma} ({\bm c}) \mathcal {U} ({\bm k}, {\bm c},s) \mathcal{P}_{\perp}
 \overline{\xi}_{3} ({\bm c}).
 \end{equation}
Now, to keep up to order $k^{2}$, the expansion
\begin{eqnarray}
\label{ape.6}
\mathcal {U} ({\bm k}, {\bm c},s)  & \simeq & e^{s \mathcal{P}_{\perp} \Lambda ({\bm c})} + \int_{0}^{s} ds_{1}
e^{s_{1} \mathcal{P}_{\perp} \Lambda ({\bm c})} \mathcal{P}_{\perp} (-i {\bm k} \cdot {\bm c}) \nonumber \\
& & \times e^{(s-s_{1}) \mathcal{P}_{\perp} \Lambda ({\bm c})}
\end{eqnarray}
is employed to get
\begin{eqnarray}
\label{ape.7}
K_{1}(s)  \simeq \frac{2 \widetilde{V}^{2} \zeta_{0}}{N d}\, \delta_{{\bm k},-{\bm k}^{\prime}} k^{2} \int_{0}^{s} ds_{1}\, e^{\overline{\lambda}_{5} s_{1}} \int d{\bm c}\, \Sigma_{x} ({\bm c}) c_{x} \mathcal{P}_{\perp} e^{(s-s_{1}) \Lambda ({\bm c})} \overline{\xi}_{3}({\bm c}),
\end{eqnarray}
since the first term on the right hand side of Eq.\ (\ref{ape.6}) gives a vanishing contribution to $K_{1}(s)$ because of symmetry. In principle, the velocity integral on the right hand side of the above equality can be evaluated by
using the same kind of approximations considered along this paper, i.e. by treating $\Sigma_{x} c_{x} \chi (c)$ as
an eigenfunction of $\Lambda^{+} ({\bm c})$ and determining the eigenvalue in a self-consistent way. Nevertheless, the contribution given by $K_{1}(s)$ seems to be negligible as compared with the other terms retained in Eq.\ (\ref{5.24}). The argument is as follows. Due to the operator $\mathcal{P}_{\perp}$ itself, the time exponential to its right decays with eigenvalues corresponding to the kinetic, non-hydrodynamic part of the spectrum of $\Lambda$. Because of symmetry considerations, the components associated to the eigenvalue $\overline{\lambda}_{4}$ and $\overline{\lambda}_{5}$ give vanishing contributions to the integral. Then, a sensible estimation of the velocity integral seems to be
\begin{equation}
 \label{ape.7a}
\int d{\bm c}\, \Sigma_{x} ({\bm c}) c_{x} \mathcal{P}_{\perp} e^{(s-s_{1}) \Lambda ({\bm c})} \overline{\xi}_{3}({\bm c}) \simeq e^{(s-s_{1})\overline{\lambda}_{5}} \int d{\bm c}\, \Sigma_{x} ({\bm c}) c_{x} \mathcal{P}_{\perp} \overline{\xi}_{3}({\bm c}).
\end{equation}
The underlying assumption is that $\overline{\lambda}_{5}$ is an upper-bound for all the kinetic modes. Note that, in particular, this is true for $\overline{\lambda}_{4}$ as obtained above. In Figs. \ref{fig1} and \ref{fig2} the obtained approximated expression for  $K_{1}(s)$ is compared with the term proportional to $a_{2} (\alpha)$ on the right hand side of Eq.\ (\ref{5.24}) for $\alpha=0.6$ and $\alpha =0.9$. Namely, the two plotted quantities are
\begin{equation}
\label{ape.8}
A_{1} e^{s \overline{\lambda}_{5}}, \quad \quad A_{1} \equiv \frac{(d+2)(d+8) a_{2} (\alpha)}{2d}\,
\end{equation}
and
\begin{equation}
\label{ape.9}
A_{2} s e^{s \overline{\lambda}_{5}}, \quad \quad A_{2} \equiv 2 \zeta_{0} \int d{\bm c}\, \Sigma_{x}({\bm c}) c_{x} \mathcal{P}_{\perp} \overline{\xi}_{3}({\bm c}),
\end{equation}
as functions of $s$.  The latter has been evaluated using the first Sonine approximation for $\chi (c)$.
It follows from the figures that for those values of $s$ for which both quantities are not negligible, it is $|A_{2}(\alpha)| s e^{s \overline{\lambda}_{5}} \ll |A_{1} (\alpha)| e^{s \overline{\lambda}_{5}}$. It is true that the two functions cross one another for large enough values of $s$, but when this happens both are very small. For instance, for $\alpha =0.6$ the curves cross at $s \simeq 6,7$, and their value is of the order of $10^{-4}$. For $\alpha = 0.9$, the value of the functions at the intersection roughly $-10^{-12}$. Similar behaviors arre obtained for other values of $\alpha$. The above results indicate that the contribution $K_{1}(s-s^{\prime})$ to the correlation of the noise term in the energy equation can be safely neglected. The term $K_{2}(s)$ given in Eq.\ (\ref{ape.4}) can be analyzed in a similar way, reaching the same conclusion. This justifies neglecting the last term on the right hand side of Eq.\ (\ref{apd.1}).

\section{Correlation between the noise terms in the velocity and energy equations}
\label{ApF}
Taking into account that $\delta_{1}^{(I)} {\bm \phi} ({\bm k},s)$ is at least of first order in ${\bm k}$, it follows from Eqs.\ (\ref{4.4}), (\ref{5.22}) and the definition of $\widetilde{\bm W} ({\bm k},s) $ given above Eq.\ (\ref{4.15}) that to Navier-Stokes order, i.e. to second order in the gradients, it is
\begin{equation}
\label{apf.1}
\langle \widetilde{\bm W} ({\bm k},s) \widetilde{\mathcal{E}}({\bm k}^{\prime},s^{\prime}) \rangle_{H} = -i {\bm k} \cdot \langle {\sf R} ({\bm k},s) S_{\epsilon} ({\bm k}^{\prime}, s^{\prime} \rangle_{H} - \frac{2}{d}
{\bm k} \cdot \langle {\sf R} ({\bm k},s) {\bm Z}({\bm k}^{\prime},s) \rangle_{H} \cdot {\bm k}^{\prime}.
\end{equation}
Consider first
\begin{eqnarray}
\label{apf.2}
\langle R_{ij} ({\bm k},s)S_{\epsilon} ({\bm k}^{\prime},s^{\prime} \rangle_{H} &=& \frac{2}{d}\, \int_{0}^{s} ds_{1} \int d{\bm c}\ \int d{\bm c}^{\prime}\, c^{\prime 2} \Delta_{ij}({\bm c}) \mathcal{U} ({\bm k},{\bm c},s_{1}) \nonumber \\
&& \times\mathcal{P}_{\perp} ({\bm c}) \langle \widetilde{S}({\bm k},{\bm c},s-s_{1}) \widetilde{S} ({\bm k}^{\prime}, {\bm c}^{\prime},s^{\prime}) \rangle_{H} \nonumber \\
&=& \frac{2 \widetilde{V}^{2}}{N d}\, \delta_{{\bm k},-{\bm k}^{\prime}} \int_{0}^{s} ds_{1} \int d{\bm c}\ \int d{\bm c}^{\prime}\, c^{\prime 2} \Delta_{ij}({\bm c}) \mathcal{U} ({\bm k},{\bm c},s_{1}) \nonumber \\
&& \times \delta (s-s_{1}-s^{\prime}) \widetilde{\Gamma} ({\bm c},{\bm c}^{\prime}).
\end{eqnarray}
The right hand side on the above equation contains the integral
\begin{equation}
\label{apf.3}
I_{ij} \equiv \int d{\bm c} \int d{\bm c}^{\prime}\, c^{\prime 2} \Delta_{ij}({\bm c}) \mathcal{U} ({\bm k},{\bm c},s_{1}) \widetilde{\Gamma} ({\bm c},{\bm c}^{\prime}).
\end{equation}
Since there is already an explicit $k$ factor in Eq. (\ref{apf.1}), this quantity is needed to first order in $k$. Therefore, the approximation
\begin{eqnarray}
\label{apf.4}
\mathcal{U} ({\bm k},{\bm c},s_{1}) & \simeq&  e^{s_{1} \mathcal{P}_{\perp} \Lambda({\bm c}) \mathcal{P}_{\perp}}
\nonumber \\
&& - \int_{0}^{s_{1}} ds_{2}\, e^{s_{2} \mathcal{P}_{\perp} \Lambda({\bm c})
 \mathcal{P}_{\perp}} \mathcal{P}_{\perp}
i {\bm k} \cdot {\bm c}\,  e^{(s_{1}-s_{2}) \mathcal{P}_{\perp} \Lambda({\bm c}) \mathcal{P}_{\perp}}
\end{eqnarray}
is used. The zeroth order in $k$ contribution to $I_{ij}$ is
\begin{eqnarray}
\label{apf.5}
I_{ij}^{(0)} & = & \int d{\bm c} \int d{\bm c}^{\prime}\, c^{\prime 2} \Delta_{ij}({\bm c}) e^{s_{1} \mathcal{P}_{\perp} \Lambda({\bm c}) \mathcal{P}_{\perp}} \widetilde{\Gamma} ({\bm c},{\bm c}^{\prime}) \nonumber \\
& =& \int d{\bm c} \int d{\bm c}^{\prime}\, c^{\prime 2} \Delta_{ij}({\bm c})e^{s_{1}  \Lambda ({\bm c})} \widetilde{\Gamma} ({\bm c},{\bm c}^{\prime}) \nonumber \\
&=&  \int d{\bm c}\, \Delta_{ij}({\bm c}) e^{s_{1} \Lambda ({\bm c})} \int d{\bm c}^{\prime}\, c^{\prime 2} \widetilde{\Gamma} ({\bm c},{\bm c}^{\prime}) =0,
\end{eqnarray}
since $\widetilde{\Gamma} ({\bm c},{\bm c}^{\prime})$ is invariant under rotations of ${\bm c}$ and ${\bm c}^{\prime}$.  The first order in $k$ contribution to $I_{ij}$ is
\begin{equation}
\label{apf.6}
I_{ij}^{(1)} = - i {\bm k} \cdot \int_{0}^{s_{1}} ds_{2}\, \int d{\bm c}\, \Delta_{ij}({\bm c}) e^{s_{2} \Lambda
({\bm c})} {\bm c} \mathcal{P}_{\perp} ({\bm c}) e^{(s_{1}-s_{2}) \Lambda ({\bm c})} \int d{\bm c}^{\prime}\, c^{\prime 2}
\widetilde{\Gamma} ({\bm c},{\bm c}^{\prime}).
\end{equation}
It is
\begin{equation}
\label{apf.7}
\langle \overline{\bm \xi}_{2}({\bm c}) | e^{s \Lambda ({\bm c})} \int d{\bm c}^{\prime}\, c^{\prime 2} \widetilde{\Gamma} ({\bm c},{\bm c}^{\prime}) \rangle =0
\end{equation}
and also
\begin{equation}
\label{apf.8}
\int d{\bm c}\, \Delta_{ij}({\bm c}) e^{s \Lambda ({\bm c})} {\bm c} \overline{\xi}_{\beta}({\bm c}) =0,
\end{equation}
for $\beta=1,3$. Therefore, the operator $\mathcal{P}_{\perp}({\bm c})$ on the right hand side of Eq.\ (\ref{apf.6}) can be omitted, and the expression is seen to vanish since it has the form
\begin{equation}
\label{apf.9}
I_{ij}^{(1)} = - i {\bm k} \cdot \int_{0}^{s_{1}} ds_{2}\, \int d{\bm c}\, \Delta_{ij}({\bm c}) e^{s_{2} \Lambda
({\bm c})} {\bm c} g(|{\bm c}|,s_{1}-s_{2})=0.
\end{equation}
To compute the second term on the right hand side of Eq.\ (\ref{apf.1}) to Navier-Stokes ($k^{2}$) order, the correlation function appearing there must be evaluated to order zero. Then the projection operators can be eliminated by using the same kind or arguments as above and it is obtained
\begin{eqnarray}
\label{apf.10}
\langle R_{ij} ({\bm k},s) Z_{l}({\bm k}^{\prime},s) \rangle_{H}  &\simeq &\frac{\widetilde{V}^{2}}{N}\, \delta_{{\bm k}, -{\bm k}^{\prime}} \int_{0}^{s}ds_{1} \int_{0}^{s^{\prime}} ds_{2}\, \delta (s-s_{1}-s^{\prime}+s_{2}) \nonumber \\
& & \times \int d{\bm c} \int d{\bm c}^{\prime} \Delta_{ij} ({\bm c}) \Sigma_{l} ({\bm c}^{\prime}) e^{s_{1} \Lambda ({\bm c})}
e^{s_{2} \Lambda ({\bm c}^{\prime})} \widetilde{\Gamma} ({\bm c},{\bm c}^{\prime})=0,
\end{eqnarray}
again because of symmetry considerations as a consequence of the isotropy of $\Lambda$ and the invariance of $\Gamma ({\bm c},{\bm c}^{\prime})$ under rotations. This completes that proof that the correlation in Eq.\ (\ref{apf.1}) vanishes.

\end{document}